# Interface superconductivity: History, development and prospects

Juan Pereiro[a,b], Alexander Petrovic[a], Christos Panagopoulos[a] and Ivan Božović[b]*

[a]Division of Physics and Applied Physics, School of Physical and Mathematical Sciences, Nanyang Technological University, 21 Nanyang Link, 637371 Singapore

[b]Condensed Matter Physics and Material Science Department, Brookhaven National Laboratory, Upton, New York 11973-5000, USA

## Abstract:

The concept of interface superconductivity was introduced over 50 years ago. Some of the greatest physicists of that time wondered whether a quasi-two-dimensional (2D) superconductor can actually exist, what are the peculiarities of 2D superconductivity, and how does the reduced dimensionality affect the critical temperature (Tc). The discovery of high-temperature superconductors, which are composed of coupled 2D superconducting layers, further increased the interest in reduced dimensionality structures. In parallel, the advances in experimental techniques made it possible to grow epitaxial 2D structures with atomically flat surfaces and interfaces, enabling some of the experiments that were proposed decades ago to be performed finally. Now we know that interface superconductivity can occur at the junction of two different materials (metals, insulators, semiconductors). This phenomenon is being explored intensely; it is also exploited as a means to increase Tc or to study quantum critical phenomena. This research may or may not produce a superconductor with a higher Tc or a useful superconducting electronic device but it will likely bring in new insights into the physics underlying high-temperature superconductivity.

Keywords: Interface, Superconductivity, Field effect, Cuprates, High Temperature Superconductors, 2D.

*Corresponding author: Condensed Matter Physics and Materials Science Department, Building 480, Brookhaven National Laboratory, Upton, New York, 11973-5000, USA. E-mail: bozovic@bnl.gov. Phone number: (631) 344 4973. Fax number: (631) 344 4071.

# 1. Introduction

In the last few years, it was discovered that interface superconductivity appears in some insulating materials ($SrTiO_3$, $La_2CuO_4$) once they are brought into contact with suitable (but also non-superconducting) partner compounds, with or even without an applied bias voltage. Our primary goal here is to present in one place a readable and updated overview of these exciting new results, primarily for novices or researchers in other fields. We will try also to provide some context and outlook, indicating why these advances may prove to be important.

The key distinguishing feature here is that the induced superconductivity may be considered as 2D. Whether and when such a categorization is justified is somewhat vague and dependent on circumstances. A reasonable criterion is that the thickness $d$ of the layer were superconductivity occurs is comparable to or smaller than the superconducting coherence length $\xi$ - which, of course, is dependent not just on the material but also on the temperature.

The question is whether it is possible, and how, to physically realize a 2D superconductor. Almost 50 years ago, Saint-James and De Gennes [1] suggested that in an ideal sample with a homogeneous order parameter, when the applied magnetic field is decreased continuously from a higher value towards the critical field, nucleation of superconducting regions should always occur near the surface of the sample. In 1964 Strongin *et al*. measured the critical field at which superconductivity nucleates on the surface of Pb doped with Bi [2]. The same year Ginzburg published the article “On surface superconductivity” [3]. He proposed two different situations whereby this could happen: (i) the electrons in surface states fill partially the surface bands, and hence in an otherwise insulating material a near-surface layer can acquire metallic character and becomes superconducting, and (ii) attraction between carriers occurs only near the surface due to surface phonons, variation in screening, etc., while in the bulk the interaction is repulsive. He even envisioned that the electron interaction may be modified and controlled by depositing “dielectric or monomolecular layers on the surface” [4].

In 1968, the first observation of enhanced superconductivity in heterostructures was reported [5]; $T_c$ of Al increases in superlattices where Al alternates with Cu or Sn. However, it took several decades before the technology necessary to put Ginzburg’s ideas to proper test was fully developed. In order to obtain interface superconductivity, the ability of fabricating 2D structures is needed. In the last two decades, great strides have been made in this direction, with the development of thin-film deposition techniques controlled on the scale of molecular and atomic layers such as molecular beam epitaxy [6] and pulsed lased deposition [7-11]. Today, thin films of a range of compounds, including a variety of complex oxides, can be synthesized with atomically flat surfaces. Moreover, atomic-layer engineering is possible, and the capability to fabricate high-quality multilayers and superlattices with atomically flat interfaces, and with constituent layers down to one unit cell (1 UC) thick, has been amply demonstrated. Thus, the technical requirements to obtain 2D superconductivity have been reached.

In parallel, solid-state chemists synthesized a range of ‘natural superlattices’ – layered compounds in which e.g. metallic layers, just one or few atoms thick, alternate with insulating layers – and some of these turned to be superconducting. The material with the highest known $T_c$ (135 K at normal pressure and 165 K under high pressure), $HgBa_2Ca_2Cu_3O_{8+\delta}$, belongs to this structural type [12]. In fact, all the compounds with $T_c$ higher than about 30 K, including cuprates, pnictides, $MgB_2$, etc., happen to be layered. This fact has, of course, greatly invigorated the interest in the effects of reduced dimensionality on superconductivity.

It has been realized since long ago that 2D is special for superconductivity. The Mermin-Wagner-Hohenberg theorem asserts that in 1D or 2D systems true long range order is destroyed by thermal fluctuations at any nonzero temperature $T > 0$. [13-15]. However, Berezinskii, Kosterlitz and Thouless argued that in 2D a quasi-long-range order and in particular superconductivity may occur [16-18]. Hence, 2D is the lower critical dimension for superconductivity, as it is for localization, leading to interesting effects of competition between the two. 2D is also special for continuous phase transitions, since in this case the length scale drops out for resistance and its value at the critical point should be universal. However, in order to keep the size of this article reasonable, we will refrain from elaborating further on the theory on 2D superconductivity and instead refer the reader to several excellent reviews of this topic [19-21].

For the same reason, we will omit discussion of proximity effect between a superconductor (*S*) and a normal metal (*N*), although one could imagine near the *S*-*N* interface a distinct layer, in some cases very thin, in which the superfluid density ($n_s$) and $T_c$ are reduced compared to those found in *S*. Much has been written on this subject starting with the pioneering work by De Gennes and Deutscher [22] and including several excellent reviews [23,24]. Here we have very little to add, except perhaps that in the case of cuprates which show high-temperature superconductivity (HTS), the proximity effect seems to be quite anomalous. If an optimally doped HTS layer is placed next to a less doped layer of the same cuprate, one of the following two things happens. If the latter is doped to such a low level that it is insulating, the superconducting order parameter decays on a very short length scale, 1 Å; one could say that there is no proximity effect whatsoever [25]. For a slightly higher doping, when the second layer becomes metallic and superconducting (although still very underdoped and with a very low $T_c$), suddenly the proximity effect shoots up to become anomalously long-ranged, at least two orders of magnitude longer than what one would have expected based on the standard De Gennes-Deutscher theory; in reference to this fact, this is sometimes called the Giant Proximity Effect [26,27]. We feel that this phenomenon is quite interesting since it is one of very few properties of the *superconducting* state in cuprates that appears strange and anomalous, but it also falls outside of the scope of this review.

The structure of the article is as follows. In Section 2, we discuss superconductivity occurring at the interface between two non-superconducting materials. In Section 3, we review enhancement of superconductivity at the interface between two compounds, one or both of which are superconducting themselves. In Section 4, we discuss surface superconductivity induced by chemical doping or by external electric field effect. In the final Section 5, we offer few concluding remarks and briefly mention some proposals and ideas for future research.

## 2. Interface superconductivity

We will use the term 'interface superconductivity' primarily to describe the situation where superconductivity occurs at or near the interface between two materials neither of which is superconducting by itself when it is in single-phase bulk form, or when it is deposited as a single-layer film. This type of interface superconductivity has been observed in very different material families and, as we will see in this section, the causes that trigger the appearance of superconductivity are also diverse – dislocations, chemical doping, charge transfer, etc.

One can talk of interface superconductivity also in a different situation, namely when one or both materials in the heterostructure are superconducting, but when a thin interfacial layer is formed with $T_c$ significantly higher than in the bulk of either constituents. However, this second case is more difficult to detect and study experimentally, since most standard methods for characterizing the superconducting state do not have the necessary nm-scale spatial resolution. The exception is the (so far, rare) situation when the interface layer has $T_c$ that is higher than the *maximum* value that could be attained in either of the two materials at *any* level of chemical doping. To differentiate the two cases, we will refer to this situation as *interface-enhanced superconductivity*, and discuss it in greater detail in Section 3.

### 2.1 Semiconductor/semiconductor interfaces: chalcogenides

The first indications of interface superconductivity in chalcogenides were reported almost four decades ago [28,29]. Searching for excitonic superconductivity, Strongin *et al*. found that PbTe layer coated with indium or lead showed $T_c \approx 6$ K; this comparatively high $T_c$ remained basically unexplained. [28,29]. More recently, indications of superconductivity in superlattices of chalcogenide semiconductors (PbTe/SnTe, PbTe/$Pb_{1-x}Sn_xTe$, PbSe/PbS, PbTe/PbSe, PbS/PbSe, PbTe/YbSe, PbTe/YbS, PbSe/EuS) have been reported by several groups [30-32] These compounds crystallize in a simple cubic lattice of NaCl type. Most of them are narrow bandgap ($E_g$) semiconductors: $E_g$ = 0.165 eV in PbSe, $E_g$ = 0.190 eV in PbTe, and $E_g$ = 0.286 eV in PbS, respectively. YbS, YbSe and EuS are wide bandgap semiconductors with $E_g$ > 1.5 eV. It is worth noting that all these compounds show anomalously strong dependence of the energy gap on pressure: $dE_g/dP = -9.1\times10^{-3}$ eV/Kbar in PbSe, $-7.4\times10^{-3}$ eV/Kbar in PbTe and $-9.15\times10^{-3}$ eV/Kbar in PbS, respectively.

In the experiments of Murase *et al*. [31], chalcogenide superlattices were grown by vapor deposition on (001) KCl and (111) $BaF_2$ substrates. Subsequent studies by Mironov *et al*. [32] and by Fogel *et al*. [30] employed thermal evaporation on (001) KCl and mica substrates. All the heterostructures that showed superconductivity contained a narrow bandgap semiconductor. The lattice mismatch between different compounds varies from 2% between PbTe and SnTe to 13% in the PbTe/YbS system. The superlattices are unstable due to the accumulated strain and tend to degrade with time when subjected to thermal cycles; in consequence, the lifetime during which the samples remain continuous and conducting is very limited. As a general rule, the stability increases if the number of superlattice periods is low since less strain is accumulated. Edge misfit dislocation (EMD) grids have been observed (Figure 1) at the interface between the different layers for samples grown on KCl. The period of the grids ($D_g$) depend on the mismatch ($f$) and on the modulus of the Burgers vector of dislocations ($b$) according to the relationship $D_g = b/f$. When (111) $BaF_2$ or mica substrates are used, the EMD grids appear to be incomplete or do not appear whatsoever.

Note that PbTe, PbSe, PbS, SnTe, and GeTe can be made superconducting also by heavy chemical doping with acceptors [33]. However, in each case $T_c < 1$ K, while in the superlattices discussed here $T_c$ is much higher, between 2 K and 6 K. Given the high $T_c$ and the fact that it correlated with the thickness of PbTe layers, Murase *et al.* [31] concluded that superconductivity originates from postulated lead precipitates. However, two years later Mironov *et al.* [32] noticed that superconductivity only occurred in samples that contained EMD grids, and concluded that the locus of superconductivity was at the interfaces.

In 1996-2008 Fogel *et al*. published a series of articles that provided most of the details we know about the superconductivity in chalcogenide superlattices [30]. Their main results can be summarized as follows. (i) All the superlattices that show superconductivity are metallic. When the temperature coefficient of resistivity is negative (semiconducting), superconductivity is not observed. (ii) The superconducting transition is incomplete if the EMD grid is not completely formed, as happens to be the case when the layers are too thin or when $BaF_2$ or mica are used as substrates. The existence of the threshold layer thickness (the minimum needed for superconductivity to appear) was attributed to the necessity of accumulating enough elastic energy to generate the complete EMD grid (Figure 2). (iii) There is a correlation between $T_c$ and the period of the EMD grid; for $D_g > 10$ nm, a reduction of $T_c$ is observed as $D_g$ is increased. (iv) Measurements of the critical magnetic field showed a crossover from 3D to 2D behavior as the temperature was decreased. (v) The thickness of the superconducting layer calculated from the value of the anisotropy parameter was found to be between 10 nm and 22 nm. (vi) If the layers are thick enough, superconductivity can be observed in a bilayer structure, i.e., from a single interface.

These results were taken as evidence that such heterostructures contain superconducting layers located at the interfaces between two semiconductors separated by interlayers of non-superconducting material. A model was proposed in which superconductivity appears due to the band inversion that occurs in the EMD grid. Since the values for $dE_g/dP$ are anomalously large, in the narrow-bandgap compounds the strain field around the dislocations induces a reduction of the energy gap larger than the gap itself; the material becomes metallic and superconductivity occurs around the dislocations, presumably by the conventional Bardeen-Cooper-Schrieffer mechanism. While the experiments of Fogel's group apparently established a relationship between superconductivity and the EMD grid, and very probably that interface metallization occurs by gap closing or charge transfer across the interfaces, some questions remain open – most notably, why $T_c$ is so much higher in superlattices.

### 2.2. Insulator/insulator interfaces: $LaAlO_3/SrTiO_3$ and related systems

Superconductivity and insulating behavior have generally been considered as mutually exclusive phenomena, since the existence of charge carriers at the Fermi level is a prerequisite for the formation of superconducting condensate. However, groundbreaking work over the past decade on electronic reconstruction at interfaces between two insulating perovskites has led to the synthesis of superconducting heterostructures made of components that are insulating *per se*, *i.e*., in bulk or single-phase thin film form.

In 2002, Ohtomo and Hwang synthesized heterostructures combining $LaTiO_3$ (LTO), a Mott insulator, and $SrTiO_3$ (STO), a band insulator. They found that the STO/LTO bilayers were metallic [34], and this was ascribed to charge transfer across the interface – electron depletion in LTO and accumulation in STO [35]. Two years later, Ohtomo and Hwang reported metallic behavior in similar heterostructures except that LTO was replaced by $LaAlO_3$ (LAO), which itself is also a band insulator [9]. When the STO/LAO interface is formed between $(LaO)^+$ and $(TiO_2)^0$ planes, a high mobility electron gas is generated, with the Hall mobility $\mu_H$ reaching $10^4$ $cm^2V^{-1}s^{-1}$ at low temperature. Electron energy loss spectroscopy (EELS) measurements, made using a transmission electron microscope (TEM), indicated that the extent of this gas was limited to an approximately 2 nm thick layer near the interface. In contrast, hole-doped interfaces between $(AlO_2)^-$ and $(SrO)^0$ remained insulating.

A more detailed analysis of the LAO/STO system by Nakagawa *et al*. suggested that the driving force behind the electronic reconstruction at the interface is the so-called "polarization catastrophe" [36]. This is illustrated in figure 3. At the junction between a polar and a non-polar material, charge transfer must take place across the interface in order to avert any divergence of the electric potential. For the n-type $(LaO)^+/(TiO_2)^0$ interface, this results in the transfer of half an electron per unit cell to the $TiO_2$ plane. Charge compensation at the p-type $(AlO_2)^-/(SrO)^0$ interface requires injection of half a hole into the SrO plane, but due to the lack of electronic states available, this may only be accomplished by creating oxygen vacancies at the interface, i.e. via atomic rather than electronic reconstruction. The resulting compensated interface contains no free charge carriers and hence remains insulating: this is in contrast with the electron-donating role of oxygen vacancies in most complex oxides.

For n-type conducting interfaces, a critical thickness of four unit cells was soon established for the LAO layer capping STO, below which metallic behavior could not be observed [10]. Energy considerations reveal that the electric potential generated by four uncompensated LAO layers corresponds to the barrier potential for charge transfer across the interface. Nevertheless, metallic behavior may still be obtained for systems with subcritical LAO capping layers by applying a gate voltage across the interface, thus enabling the metallic conductivity of the interface to be switched on and off. Recently, tunneling spectroscopy measurements on LAO/STO structures indicated that the electrons in the channel are actually interacting and should be considered as an electron liquid [37].

In addition to the polarization catastrophe, it rapidly became apparent that oxygen doping has a significant effect on the transport properties of these interfaces. The predicted sheet carrier density $n_s$ resulting from electronic reconstruction driven by polar discontinuities is merely $3.3\times10^{14}$ $cm^{-2}$, while the measured values for $n_s$ exceeded $10^{16}$ $cm^{-2}$ for films grown at low oxygen pressure [38-40]. Herranz *et al*. [38] estimated the thickness of the conducting layer to be around 530 μm, approximately equal to the thickness of the STO substrate, likely indicating that it was loaded with oxygen vacancies. Analysis of the Shubnikov-de Haas oscillations in the magnetoresistance also indicated a 3D nature of the electron gas.

The role played by oxygen vacancies in determining the dimensionality of the interfacial electron gas was elegantly illustrated by Basletić *et al.*, who used conducting-tip atomic force microscopy in a cross-sectional configuration to measure the local conductance across LAO/STO interfaces synthesized under both low and high oxygen pressure conditions [41]. For samples grown at an oxygen pressure of $10^{-6}$ mbar, the conducting region may indeed extend over hundreds of microns (Figure 4, right). Conduction in this case is therefore clearly dominated by oxygen vacancies within the STO substrate itself. In contrast, when the film was cooled down under 300 mbar of oxygen pressure, the width of the conducting region was reduced to 7 nm (Figure 4, left). Here, the oxygen vacancies have been filled and the dominant cause of metallic conduction is presumably the electronic reconstruction required to avert the polarization catastrophe.

It is also possible to introduce oxygen vacancies via nanoscale field effect doping, using an AFM probe as the voltage gate contact [42]. By modulating the gate voltage at the AFM tip, stable conducting regions may be reversibly written onto or erased from 3 unit-cell thick LAO films on a STO substrate. This allows nanowires and tunnel junctions to be created, potentially opening a new route towards the study of single-electron effects and low-dimensionality over a broad temperature range [43]. These so-called "SketchFET" devices may be able to operate at frequencies extending up to the GHz regime, suggesting a potential application as high-performance nano-transistors.

Let us now turn to the low-temperature ground state of this interface. Superconductivity has been known to occur in doped STO for many years; in fact, niobium-doped STO was one of the first materials to be identified as having a multi-band superconducting order parameter [44]. Furthermore, it was suggested long ago that field-effect doping of STO could lead to the formation of a superconducting 2D electron gas (2DEG) at the surface [45]. Speculations were even made that HTS may be achieved because polarization in STO layers could act as the "glue" to bind surface electrons [46,47] However, the real breakthrough was the discovery by Reyren *et al*. that LAO/STO interfaces become superconducting at around 200 mK [11]. The in-plane coherence length was estimated to be $\xi \approx 70$ nm, while $n \approx 4\text{x}10^{13}$ $\text{cm}^{-2}$. This carrier density imposed an upper limit of 10 nm for the thickness of the superconducting slab, placing it firmly in the 2D regime. Further evidence for 2D superconductivity was provided by measuring the anisotropy parameter $\varepsilon \equiv H_{c2}^{//}/H_{c2}^{\perp} \approx 25$ [48] (Figure 5).

As mentioned in the Introduction, in the 2D limit the superconducting transition should belong to the well-known Berezinskii-Kosterlitz-Thouless (BKT) universality class [16,18]. Within this model, thermal excitations of the superconducting order parameter take the form of vortices and anti-vortices; below the transition temperature $T_{BKT}$ these are bound into vortex/anti-vortex pairs, creating a globally coherent ground state. Experimentally, the presence of a BKT transition may be quantitatively determined via transport measurements in two ways: first, the *I-V* curves should exhibit a power-law dependence $V \sim I^{\alpha}$, where $\alpha = 3$ at $T_{BKT}$ and diverges at lower temperature. Second, in the vicinity of the transition the resistance $(R)$ of the material should follow the relation $R = R_0 \exp(-bt^{-1/2})$, where $R_0$ is a material-dependent parameter, $b$ is related to the vortex creation energy, and $t = (T/T_{BKT} - 1)$ is the reduced temperature. As may be seen in figure 6, both of these conditions were fulfilled in the LAO/STO system, yielding $T_{BKT} \approx 190$ mK for an 8 UC thick LAO capping layer. It should be noted that $R(T)$ at the transition often displays a "hump" structure before falling to zero: this is characteristic of a BKT transition and is generally attributed to the small temperature offset between the initial formation of free vortices and their subsequent condensation into bound pairs.

A thorough theoretical investigation of the LAO/STO system was subsequently carried out by Schneider *et al.* [49]. Their scaling analysis showed that the superconducting transition is well described by the BKT model with a finite size effect, caused largely by inhomogeneity within the conducting channel. The characteristic length-scale of this inhomogeneity imposes a cut-off on the divergence of correlation length at $T_{BKT}$, resulting in a slight broadening of the transition. The observed variation of this length-scale with the gate voltage implies that field-effect doping changes not only the carrier density but also the landscape of disorder at the interface. Above the quantum critical point, the sheet conductivity $\sigma$ varies as $T^2$ characteristic of a 2D Fermi liquid, with a crossover to a $T$-linear behavior at higher temperature. For large negative gate voltages (*i.e.*, in the electron-depleted insulating phase), one observes $\sigma \sim ln(T)$ temperature dependence, expected for a weakly-localized Fermi liquid. A small systematic deviation observed at low temperature is due to the same finite-size effect responsible for broadening the BKT transition. Together, these effects create an entirely 2D phase diagram laden with novel physics. Further work has highlighted qualitative distinctions between the 2D metal-insulator transition (MIT) in LAO/STO and typical 2D MITs seen at semiconductor interfaces [50]. In particular, for sub-critical carrier densities a threshold behavior appears in the field-dependent current density $J(E)$ in LAO/STO which is absent in analogous semiconducting interfaces. There is also little temperature dependence observed for the MIT in LAO/STO, where disorder seems to play a more important role. While the critical exponents are consistent with 2D percolation in both semiconducting and LAO/STO interfaces, the exact mechanism for the MIT in each system remains an open topic.

The growth and structural properties of the LAO/STO heterostructures have recently been characterized in greater detail [51]. A thorough assessment of the effect of oxygen vacancies on the electronic properties of this system has also been carried out [52]. The key finding was that regardless of the capping layer thickness or initial oxygen pressure during film growth, vacancies may be reliably eliminated by an appropriate annealing process at high oxygen pressure.

Very recently, Eom and coworkers used pulsed-laser deposition to insert a single atomic layer of a rare-earth oxide (*R*O), where *R* is lanthanum (La), praseodymium (Pr), neodymium (Nd), samarium (Sm), or yttrium (Y), into an epitaxial STO matrix. [53]. The growth was controlled by reflection high-energy electron diffraction (RHEED), see figure 7. The structural quality of these STO-RO-STO heterostructures was superb. AFM image of the surface of a complete heterostructure (Figure 8) showed steps and terraces as seen in the substrate itself. Cross-section TEM (Figure 9) showed atomically abrupt interfaces and no observable cation interdiffusion. Electron energy-loss spectroscopy (EELS) was used to probe charge transfer from the *R*O layer to nearby Ti states, by depth profiling of the $Ti^{3+}$ to $Ti^{4+}$ ratio. The results indicated that the carriers were confined to within ~1 nm of the interface.

However, Jang *et al*. have also demonstrated that electronic charge-transfer is not the whole story [54]. Synchrotron-based X-ray diffraction (XRD) showed unit-cell doubling $TiO_6$ octahedra rotations in the $RTiO_3$ layer, which are well ordered in the interfacial plane with in-plane domain size > 60 nm, but rapidly decay into the $SrTiO_3$ matrix. In addition, epitaxial strain in the interfacial $RTiO_3$ layer also affects the interface conductivity. $LaTiO_3$, $PrTiO_3$, and $NdTiO_3$ layers at the interface are strained under biaxial compression, but $SmTiO_3$ and $YTiO_3$ layers are under bi-axial tension. These differences are likely to constitute the reason why the structures with La, Pr, and Nd ions feature a metallic 2DEG in the inserted layer, whereas structures with Sm or Y ions remain insulating.

A substantial body of work indicates that cation interdiffusion in fact plays a substantial role at such complex oxide interfaces. High-resolution Rutherford backscattering spectrometry showed that La from a capping LAO layer grown by pulsed laser deposition (PLD) is able to diffuse deep into a STO substrate (where it acts as an electron donor), with Sr moving in the opposite direction [55]. This cation mixing destroys the electric dipole in the LAO layer and results in n-type doping of the STO substrate, implying that La diffuses preferentially into the STO compared to Al (which acts as an acceptor) [56]. Such a significant degree of La interdiffusion would be expected to result in a carrier-doped "tail" stretching hundreds of Angstroms into the STO substrate. Dubroka *et al*. have indeed recently found evidence for such a long La "tail" using ellipsometry measurements [57].

Furthermore, density functional theory (DFT) calculations indicate that an interface exhibiting cation intermixing is generally thermodynamically favorable compared to an atomically abrupt interface [50]. It appears that cation interdiffusion must therefore be added to the polar catastrophe and oxygen vacancy creation as mechanisms responsible for carrier injection at the LAO/STO interface. Whether or not interfaces grown by molecular beam epitaxy (MBE) rather than PLD might exhibit a more abrupt compositional crossover (due to the lower intrinsic deposition energies) remains to be determined.

Altogether, this tour-de-force materials science research has underlined the complexity of interface physics which, apart from electronic phenomena, also includes subtle chemical and structural effects; these aspects are closely intertwined.

In parallel to the work on LAO/STO, Hwang and his coworkers fabricated STO-based heterostructures by selectively doping some layers with niobium, which is known to electron-dope STO and render it metallic. [58]. A trilayer structure, containing one layer of STO, followed by a layer of niobium-doped STO, and covered with another STO layer, was shown to behave as a highly 2D superconductor with $T_c$ = 307 mK and $\varepsilon \approx 31$. The observation of angle-dependent Shubnikov-de Haas oscillations in magnetoresistance indicated that a 2D Fermi surface has been engineered within the doped STO layer, a remarkable feature given that bulk Nb-doped STO is a 3D superconductor. In the normal state at 2 K, this layer exhibits a mobility of 1,100 $cm^2V^{-1}s^{-1}$, which is exceptionally high for such a heavily doped semiconductor (the mobility is generally limited by disorder). Furthermore, the ratio between the superconducting coherence length and the mean free path, $\xi/l = 0.21$, implies that the superconductivity is close to the clean limit. At low temperature, due to the low effective mass of the carriers, the system approaches the quantum limit $k_BT << \hbar\omega_c$ (where $\omega_c$ is the cyclotron frequency). This may open the door to some interesting new physics, such as the possibility of re-entrant superconductivity [59].

### 2.3 Metal/insulator interfaces: $La_{1.55}Sr_{0.45}CuO_4$ / $La_2CuO_4$

***Discovery.*** In 2005-2010, the Božović's group at Brookhaven National Laboratory in USA carried out a systematic investigation of interface superconductivity in various heterostructures (bilayer, trilayer, and superlattices) made with layers of overdoped, metallic $La_{2-x}Sr_xCuO_4$ with x > 0.35 and insulating $La_2CuO_4$; in this section, we will denote them as *M* and *I*, respectively. Note that neither of these is superconducting in bulk or single-phase film form, see figure 10. Nevertheless, bilayers were found to be superconducting. This discovery was first reported at a number of conferences (in 2006 at the Stripes conference in Erice, Sicily; in 2007 at the APS March Meeting, the RTS Workshop in Loen, Norway, and notably the conference in Ramat Gan, Israel) and published eventually [60]. The films were synthesized using a unique atomic-layer-by-layer molecular beam epitaxy (ALL-MBE) system that incorporates *in situ* state-of-the-art surface science tools, such as time-of-flight ion scattering and recoil spectroscopy (TOF-ISARS) and RHEED. It enables synthesis of atomically smooth films as well as multilayers with perfect interfaces [61]. The films were deposited on single-crystal $LaSrAlO_4$ substrates cut perpendicular to (001) direction. A range of characterization techniques was employed to verify that surfaces and interfaces were atomically smooth.

Surprisingly, Gozar *et al.* found that $T_c$ depended on the layer sequence, see figure 10. In *I-M* bilayers where *I* layer was deposited first onto the substrate, typically $T_c \approx 15$ K was observed. In *M-I* bilayers (*M* next to the substrate), $T_c$ was about twice higher. (The highest seen so far was $T_c$ = 38 K). This asymmetry was a surprise. If the observed interface superconductivity indeed resulted from charge transfer between *M* and *I* layers, which in turn originated from the difference in the respective electrochemical potentials ($\Delta\mu$), then the order in which the layers were deposited should not matter - $\Delta\mu$ should be symmetric with respect to the layer swap, as long as the substrate influence could be neglected. Apparently, this last assumption was not justified, even when the active interface was separated from the substrate by a buffer layer few hundred Å thick.

***Key questions.*** Like in the case of metallic conductivity and superconductivity in LAO/STO heterostructures, this discovery opened a number of questions. The first was whether this is indeed an interface effect or possibly a bulk phenomenon; more quantitatively, what is the thickness of the superconducting layer. The second was the question of underlying mechanism: electronic charge transfer across the interface, *i.e.* accumulation of holes in *I* and depletion from *M*, or simply some Sr interdiffusion between the two layers. One could imagine a gradient in local Sr content going more-or-less smoothly (or perhaps in some fractal way) from its maximal value $x_{Sr}$ = 0.45 in *M* down to $x_{Sr}$ = 0 in *I*, thus passing close to the optimal value $x_{Sr}$ = 0.15 somewhere in between. While would still be a legitimate way to create an ultrathin HTS layer, this was not the goal the researchers originally set to themselves; rather, they hoped to achieve an electronic effect. The third key question was how to explain the observed asymmetry between *I-M* and *M-I* bilayers.

Answering these questions by experiment posed substantial technical challenges. To differentiate between the two mechanisms – atomic vs. electronic - one needed to measure, with atomic resolution along the *z*-axis (perpendicular to the $CuO_2$ planes and to the interface), the profiles of local Sr concentration $x_{Sr}(z)$ and of the density of mobile carriers, $n(z)$ (The analogous and equally difficult question in the case of LAO/STO interfaces is whether superconductivity is induced in STO by chemical doping from oxygen vacancies and/or La interdiffusion, or else by electron charge transfer). To fully understand this phenomenon of interface superconductivity, one would also need to know the local crystallographic and electronic structure. But the techniques for full chemical, crystallographic, and electronic characterization of buried interfaces with atomic resolution were not yet available; it was necessary to develop new techniques or at least to push the existing ones beyond the existing boundaries. The demand to measure with atomic resolution the vertical profile of the superfluid density, $n_s(z)$, appeared the most difficult, since the traditional techniques were typically limited by much larger length scales such as the in-plane coherence length (1-2 nm) or the London penetration depth (> 200 nm).

***Thickness dependence.*** Already in the first paper in this series, Gozar *et al*. [60] demonstrated that superconductivity was confined to an interfacial layer not more than 1-2 UC thick by synthesizing a series of *M-I* and *I-M* heterostructures with thick bottom layers (≥ 30 UC) while the top layer thickness was varied systematically in 0.5 UC increments. They found that $T_c$ increased with the top layer thickness until it saturated for 2 UC and thicker layers, see figure 11, and concluded that the HTS layer thickness must be restricted to just few $CuO_2$ planes.

***Chemical composition profile.*** The first evidence against massive Sr interdiffusion was obtained already in real time, during the film growth, by observing the RHEED patterns [60]. Pronounced oscillations of the intensity of the specular spot in RHEED were observed, indicating atomically smooth layer-by-layer growth. Note that every compound has characteristic form factors determined by the chemical composition, the nature of the surface states, etc., and hence it displays its own characteristic pattern of amplitude and shape of RHEED oscillations. Switching between *I* and *M*, the pattern (both the amplitude and the shape) of oscillations was observed to change abruptly on the 0.5 UC scale from the one typical of single-phase *I* films to the one typical of single-phase *M*. This indicated that interfaces were atomically sharp with respect to the cation composition, irrespective of the deposition sequence.

The next tool employed *in-situ* was TOF-ISARS [62]. Here, a beam of monochromatic (10 keV) $K^+$ ions is chopped into bunches which imping on the film surface, ejecting Sr, La, Cu, and O ions from the film, which are mass-analyzed. Figure 12 show the time evolution of the Sr peak during deposition of *I-M* and *M-I* bilayers. Taking into account that some ions are ejected from second and even deeper atomic monolayers, one gets an upper limit of 1 UC for significant Sr diffusion – probably an overestimate. This result was confirmed subsequently using EELS in a scanning transmission electron microscope (STEM), see Figure 13. Chemical interdiffusion at the interfaces was quantified by studying the Lanthanum-$M_{4,5}$ EELS edges as a function of position. The *rms* interface roughness, from the La profile, was 1.2±0.4 nm (~ 1 UC) at the *M-I* interface, which sets an upper limit to any cation intermixing. (Notice that this is also an overestimate, because it includes contributions from interface roughness, terrace steps, local variations in the substrate termination layer, etc.) The same result was obtained by analyzing Oxygen-K edge fine structure. These and other experiments [60] set an upper limit on possible cation interdiffusion to less than 1 UC and proved that interface superconductivity cannot be attributed to just cation mixing.

***Resonant soft X-ray scattering: charge density profile.*** The charge redistribution that occurs between *I* and *M* (here, *M* = $La_{1.64}Sr_{0.36}CuO_4$) layers was probed directly [63] using resonant soft X-ray scattering (RSXR*S*). The sample under study was a superlattice containing 15 periods, each consisting of 1 UC of *I* and 2 UC of *M*. None of the constituents were superconducting, but the superlattice showed a superconducting transition with $T_c$ = 38 *K*. The analysis of the structure factors from the results obtained in the RSXRS measurements produced the results summarized in Figure 14. The actual structure of the superlattice repeat unit, which contains 3 nonequivalent $CuO_2$ layers - two belonging to the central *M* layers, two belonging to the outer *M* layers, and two belonging to the *I* layers - is shown in figure 14b. The hole concentration corresponding to each of these nonequivalent layers is denoted as $p^0_{max}$, $p^0_{int}$ and $p^0_{min}$ respectively, as indicated in figure 14. The charge conservation, the scattering near the first Bragg peak ($L = 1$) and the scattering near the $L=2$ peak of the superlattice at different energies (near the La M5 edge and near the mobile carrier peak resonance), provided the three independent measurements that were needed in order to solve for $p^0_{max}$, $p^0_{int}$ and $p^0_{min}$. In the detailed analysis of the RSXRS results, Smadici *et al.* also accounted for the effect of interface roughness. In figure 14a, the solid squares show the layer-resolved hole count in an ideal structure, without accounting for the effect of the roughness; the hole count that would be obtained from the structural roughness only is represented by open circles; and the nominal distribution of $Sr^{2+}$ ions by open squares. Figure14c shows the final results. It indicates that the holes are not bound to the $Sr^{2+}$ ions, but rearrange between the layers. The electron redistribution generates the hole concentration of about 0.15 holes/Cu in the $La_2CuO_4$ layers, which are then optimally doped and sustain HTS.

***Crystallography: Madelung strain***. A detailed high-resolution XRD study of *M-I* and *I-M* bilayers, as well as *M-S* bilayers discussed later in Sect. 3, revealed an anomaly in the behavior of the lattice parameters in these heterostructures [64]. When heteroepitaxy is carried out, one usually observes that the first layers grow pseudomorphically i.e., the in-plane lattice parameters of the epitaxial layer match those of the substrate. Thus, the top layer grows strained and accumulates elastic energy. Once this elastic energy reaches a certain threshold, the film tends to relax and reduce the strain by generating defects (misfit dislocations), and the lattice parameters revert to their bulk values. During this process the volume of the unit cell is generally conserved (the Poisson Law), so when the in-plane lattice parameters are compressed, the out-of-plane lattice constant expands. However this is *not* what is observed in the *M-I* and *I-M* bilayer structures discussed here. Instead, in either case the top layer grows pseudomorphically on the bottom layer, but the out-of-plane lattice constant also matches that of the bottom layer, in gross violation of the Poisson Law. Based on extensive numerical simulations [64], it was concluded that this anomaly originates from long-range Coulomb interaction (“Madelung strain”, coming from the substrate) which along the c-axis is not completely screened.

Comparing the XRD and transport data, Butko *et al.* showed that $T_c$ of the bilayers scaled linearly with the out-of-plane lattice constant ($c_0$), see Figure 15. Such linear dependence was already noticed earlier for the whole range of Sr content in single-phase $La_{2-x}Sr_xCuO_4$ films by Sato *et al*. [65], and across various cuprate families by Locquet *et al.* [66]. The important difference here is that the expansion or compression of $c_0$ does not originate from chemical substitution of ions with different radii, but rather from epitaxial strain. At phenomenological level, one could take that this answers one of the main questions posed at the beginning of this section, *i.e*. it explains the asymmetry between *I-M* and *M-I* bilayers. The cause is structural; the crystallographic differences cause the variation in $T_c$ because of its dependence on $c_0$. This leaves open the question why $T_c$ depends on $c_0$ to begin with [67-71]. However, to answer this question one needs to know the mechanism of HTS.

***COBRA phase-retrieval crystallography: atomic displacements***. Traditional XRD provides very precise values of lattice constants but the accurate positions of atoms inside the unit cell are uncertain because of the phase problem.

More detailed information can be obtained by phase-retrieval techniques, such as Coherent Bragg-rod analysis (COBRA), which enable one to determine the complex structure factors and, thus, the full 3D electron density of the film [72-74]. In this synchrotron-based technique, X-rays impinge onto the sample at a shallow angle, which makes it sensitive to the first few UC closest to the surface; the applicability is thus limited to very thin (few UC thick) films and the surface needs to be atomically smooth. Fortunately, ALL-MBE grown bilayers fully meet these requirements, and very high quality COBRA electron density maps were obtained for single-phase optimally doped $La_{1.84}Sr_{0.16}CuO_4$, overdoped metallic LSCO, and *M-I* bilayer structures [63]. The key new finding was that in *M-I* bilayers the distance between the copper and apical oxygen atoms increases dramatically (by nearly 0.5 Å) towards the surface of the *I* layer, whereas in single-phase layers it remains constant. This phenomenon was also attributed to strong long-range Coulomb interactions in these layers.

The potential importance of this observation stems from the fact, already known for some time from a number of studies [67,68,70,71,75], that the position of apical oxygen strongly affects superconductivity in cuprates. The reasons for this are still debated [76], but it is likely that it affects the width of the relevant Cu3d-O2p electron band and hence the electron density of states. Whatever the mechanism, this result is encouraging since it points to a way of engineering cuprates to enhance $T_c$. According to Pavarini *et al*. [68], the apical oxygen distance that was observed near the surface in *M-I* bilayers should correspond to $T_c \approx$ 80-90 K, twice higher than what has been observed in LSCO so far. Unfortunately, in these *M-I* bilayers the $CuO_2$ planes with long Cu-apical O distance were undoped and insulating, even with the hole redistribution by charge-transfer taken into account. But if one finds the way of optimally doping those layers, one might end up doubling $T_c$ of the structure.

***Zn δ-doping tomography: superfluid density profile.*** Leveraging on the atomic-layer-by-layer deposition capability, Logvenov *et al.* [77] were able to carry out a simple yet compelling experiment and prove that in *M-I* bilayers HTS is confined to a single $CuO_2$ layer. The idea is similar to so-called *δ*-doping used frequently in semiconductor physics to achieve high-mobility 2D electron gas. In HTS physics, it was also used extensively to fabricate trilayer (sandwich) Josephson junctions and engineer the barrier properties. [78]. Zinc was chosen as a suitable dopant because its effect on HTS has been widely studied and it is established that even a small concentration of Zn, which substitutes for Cu, suppresses very efficiently both $T_c$ and $n_s$, without affecting the carrier density [79-82]. Logvenov *et al*. showed experimentally that replacing 3% of Cu by Zn doping reduces $T_c$ by a factor of 2 in both $La_{1.85}Sr_{0.15}CuO_4$ and $La_2CuO_{4+\delta}$ thin films. Using ALL-MBE, this doping can be confined to a single, specified $CuO_2$ layer.

The experiment consisted of synthesizing a series of *M-I* bilayers with each *M* and *I* layer exactly three UC thick (six molecular layers), the grand total of 12 $CuO_2$ planes. Except for few undoped control samples, in each film one specified $CuO_2$ layer was *δ*-doped with 3% of Zn. The position of this *δ*-doped $CuO_2$ plane, relative to the interface, was varied systematically from the $CuO_2$ plane closest to the $LaSrAlO_4$ substrate (*N*=-6) to the one nearest to the surface (*N*=6), see figure 16a. Both transport and magnetic (mutual inductance) measurements showed $T_c$ = 32±4 K in all bilayers except for the ones doped at *N = 2*. All such films had $T_c \approx$ 18 K (Figure16b). The superfluid density, also measured by mutual inductance, was reduced by a factor of 4-5 in $N = 2$ films compared to all others. This lead to the conclusion that in *M-I* bilayers superconductivity with $T_c > 30$ K occurs in a single $CuO_2$ plane, the one located in the second $CuO_2$ plane of the *I* layer counting from the nominal geometric interface with the *M* layer.

***Theoretical modeling.*** A simple model calculation [83], using the measured Thomas-Fermi screening length $\lambda_{TF} = 6$ Å [63], indicated that the locus of the nearly-optimal doping and the highest $T_c$ should in fact be in the $N = 1$ $CuO_2$ layer, if Sr concentration profile is assumed to be a perfectly sharp step-function. More realistically, if the actual Sr doping distribution as measured by TOF-ISARS and HRTEM-EELS is taken into account, a broader hole density profile $n(z)$ is obtained and the maximum $T_c$ shifts to the second $CuO_2$ layer, as observed. In other words, it appears that charge transfer and some Sr interdiffusion both occur.

More detailed calculations, and an interesting suggestion for future research, were presented by Loktev and Pogorelov [84,85]. They assumed that the collective electronic states are a superposition of almost uncoupled planar states in each $CuO_2$ plane formed by fast in-plane hopping. The role of the out-of-plane hopping, much slower than the in-plane, is reduced to establishing the common Fermi level for all layers. Using this alternative to the Thomas-Fermi treatment, they theoretically analyzed various heterostructures to be made by atomic-layer engineering and $\delta$-doping. In a structure consisting of 5 cuprate layers with the doping levels specified as $x_1 = x_5 = 0.45$, $x_2 = x_3 = x_4 = 0$, the hole distribution would be $p_1 = p_5 = 0.29$, $p_2 = p_4 = 0.12$, $p_3 = 0.08$; since $T_c$ in the layer $N = 3$ would be lower than in the layers $N = 2$ and $N = 4$, within a certain temperature range this heterostructure would behave as a superconductor-normal metal-superconductor (SNS) Josephson junction.

### 3. Interface-enhanced superconductivity

In Section 2 we have seen that if there is a "parent" compound (such as $SrTiO_3$, $La_2CuO_4$, etc.) that is not superconducting *per se* – be it insulating, semiconducting or metallic – but which can be doped to become superconducting, reaching its highest value $T_c^{max}$ at some "optimal" doping level, then interface superconductivity may be induced by charge transfer between this parent compound and a suitable partner material. Basically, the difference in chemical potentials between the two proximal materials causes charge carrier depletion and accumulation, and under well-chosen circumstances, one can achieve near-optimal doping in a thin interfacial layer. There is little mystery here - as long as $T_c^{max}$ is not exceeded. In fact, in such bilayers one would expect $T_c$ to be lower than $T_c^{max}$, for any out of a number of reasons: surface roughness, ion interdiffusion, strain arising from the lattice constant mismatch between the two compounds, atomic or electronic reconstruction (such as might arise to avoid the polarization catastrophe), 2D fluctuations, etc.

***Theoretical proposals for interface-enhanced*** $\boldsymbol{T_c}$***.*** A more exciting question is whether and under which circumstances $T_c$ in the interfacial layer could exceed $T_c^{max}$. The simplest argument why this may be possible is "doping without disorder". Note that chemical doping – in the case of $SrTiO_3$, doping by La, Nb, or oxygen vacancies, and in the case of $La_2CuO_4$ by Sr, Ba or excess (interstitial) oxygen – always increases disorder, by local distortions around the dopant ions that not only represent charge defects but also have different ionic radii and distort the lattice locally.

Such defects may act to reduce $T_c$ by localizing electrons and reducing the density of states at the Fermi level, by breaking Cooper pairs, etc. Doping by purely electronic charge transfer should not generate such defects, and $T_c$ may increase. An analogous phenomenon is the observed large increase of the Curie temperature in artificial, MBE-grown $LaMnO_3/SrMnO_3$ superlattices compared to the maximum achieved in $La_{1-x}Sr_xMnO_3$ solid solutions [86].

More sophisticate mechanisms are also conceivable. As mentioned in the Introduction, V. L. Ginzburg suggested long ago that interface enhancement of $T_c$, perhaps even room-temperature superconductivity, could be achieved by manipulating electronic and phonon surface states, the density of charge carriers, screening, the effective pairing interaction, etc. [3,4]. More recently, S. Kivelson [87] proposed that $T_c$ enhancement could be achieved by a proximity effect between a material with strong pairing but low phase stiffness (as may be the case in underdoped cuprates) and a material with weaker pairing but much higher phase stiffness (supposed to be the case in overdoped cuprates). The proximity effect could enhance the coherence in the underdoped material by quenching the strong phase fluctuations, and thus pushing the material closer to its full potential implied by the large pairing energy. Other proposals have been put forward, as well [88,89].

We have already presented in Section 2.1 one example, the artificial chalcogenide superlattices, where $T_c$ is significantly enhanced over that seen in single-phase samples. A second example may be a $YBa_2Cu_3O_7$ layer coated with Ag [90]. However, it is still unclear whether any, and which, of the above enhancement mechanisms is operative in these two cases. One more, well-documented example is presented next.

### 3.1 $La_{1-x}Sr_xCuO_4$-based *M-S* and *S'-S* heterostructure

Probably the first recorded observation of high-temperature interface superconductivity was made the Bozovic's group at Oxxel GmbH in Bremen, Germany in 2000; it was reported at several conferences first and then published a little later [61]. In figure 17, reproduced from Ref. [61], we show the *R(T)* characteristics of an *S'-S* bilayer grown by ALL-MBE, where *S'* = $La_{1.85}Sr_{0.15}CuO_4$ and *S* = $La_2CuO_{4+\delta}$. A sharp superconducting transition is seen with $T_c$ = 51.5 K, about 10 K higher than what was observed by the same group in single-phase *S* or *S'* films. However, the emphasis of that paper was on a different issue (the effect on $T_c$ of the epitaxial strain vs. interstitial oxygen incorporation), so this potentially important result passed largely unnoticed. Moreover, the actual thickness of the highest-$T_c$ layer was not determined at that time.

The next step was made by Gozar *et al*., [60] who apart from *I-M* and *M-I* bilayers also reported results on *M-S* structures, in which they observed $T_c$ = 50-52 K. The layer with this highest $T_c$ was estimated to be just 1 UC thick, while the remaining part of the S layer had $T_c$ < 40 K. This estimate was made based on the measurement of the temperature dependence of the critical current density ($j_c$). In figure 18 we reproduce $j_c(T)$ determined from two-coil mutual inductance measurements in single-phase *S* films and in *M-S* bilayers. In *S* films (red solid diamonds in figure 18), $j_c$ is linear in temperature until very near (≈ 1.5 K) to $T_c$. Such temperature dependence is expected theoretically and observed experimentally in homogeneous cuprate samples. In contrast, in *M-S* samples one can see two approximately linear regions with very different slopes and a clear break between the two at $T \approx 40$ K. This is what one would expect from two superconducting sheets with different thickness and critical temperature, say $d_1$, $T_{c1}$ and $d_2$, $T_{c2}$, respectively. The breakdown into two such components (the dashed lines in figure 18) provides $T_{c1}$ ~ 40 K (the maximum obtained in single-phase *S* films), $T_{c2}$ ~ 50 K (the same as measured in the *M-S* bilayers), and the low-temperature extrapolation of the critical current gives $d_1/d_2$ ~ 20. Since the total number of layers deposited was $d_1 + d_2$ = 20 UC, one obtains $d_2$ ~ 1 UC.

Interestingly, Gozar *et al.* were able to perform similar mutual inductance measurements on exactly the same *S'-S* bilayer sample that was studied in [61]. The film was preserved for seven years, measured again, and found not to have deteriorated. The new $j_c(T)$ data appeared quite similar to the *M-S* case shown in figure 18, demonstrating that the enhanced $T_c$ reported previously in [61] was also an interface effect.

A detailed high-resolution XRD study by Butko *et al.* [64] showed that the *c*-axis lattice constant in both the *M-S* bilayers studied by Gozar *et al.* and in the earlier *S'-S* bilayer studied in [61] was significantly elongated ($c_0 \approx 13.29$ Å) compared to what is seen in single-phase *S* or *S'* samples ($c_0 \approx 13.22$ Å), see figure 15. Through the observed empirical linear relation between $c_0$ and $T_c$, this indeed accounts for the enhanced $T_c$ – and while admittedly this relation may not be understood, the data clearly suggest that the primary cause of this $T_c$ enhancement is also structural. On the other hand, this is probably not the whole story, since the entire top *S'* layer has the same expanded $c_0$. Going for the simplest possible explanation, one may suspect that the oxygen doping is different near the interface and further out, perhaps because of modified electrostatics and strong built-in local electric field at the interface. But this suggestion has yet to be tested and verified (or refuted) experimentally.

A similar experiment, with very different conclusion, was carried out by Yuli *et al.* [91]. A series of bilayer films were grown on (100) $SrTiO_3$ substrates by pulsed laser deposition. The layer next to the substrate was a 90 nm-thick slab of $La_{2-x}Sr_xCuO_4$ with $x$ = 0.06, 0.08, 0.10, 0.12, 0.15, and 0.18. This was covered with a 10 nm-thick layer of overdoped $La_{1.65}Sr_{0.35}CuO_4$ layer. The $R(T)$ characteristics of these bilayers are shown in figure 19. An increase of $T_c$ is seen in bilayers with the underdoped bottom layer compared to the bare layer of the same composition, while no increase was seen for $x$ = 0.18. When the order of the layers was inverted, the results remained the same. Yuli *et al.* invoked neither carrier redistribution, nor Sr interdiffusion, nor structural factors, but rather concluded that the metallic cap layer enhances the phase stiffness in the underdoped bottom layer, as proposed earlier by Kivelson [87]. This interpretation was elaborated further by Berg *et al*. and Goren *et al*. [92,93]. However some questions were left open on the experimental side. To begin with, it is unclear why even the nominally optimally ($x$ = 0.15) doped film had $T_c \approx 24$ K, grossly reduced compared to $T_c^{max}$ reported by other groups, so $T_c$ enhancement here really meant $T_c$ being less reduced. Another caveat is that, judging from the experiments of the Bozovic's group, high-$T_c$ interface superconductivity occurs within a single $CuO_2$ plane, and hence atomically flat films are required in order for this plane to be continuous – which was definitely not the case here. Indeed, Yuli *et al.* did not observe any diamagnetic signal at temperature above the lower $T_c$ (that of the bottom layer). Finally, Koren *et al*. repeated the experiment [91,94] but did not reproduce the increase of $T_c$ reported by Yuli *et al*. However, they observed an increase of $T_c$ by 1.4 K in $La_{1.875}Ba_{0.125}CuO_4$ / $La_{1.65}Sr_{0.35}CuO_4$ bilayers, and again suggested that it originated in the improvement of the phase stiffness. Rout and Boudhani also relied on Kivelson's proposal to account for the observation of three transition temperatures by mutual inductance measurements in $La_{1.48}Nd_{0.4}Sr_{0.12}CuO_4/La_{1.84}Sr_{0.16}CuO_4$ bilayers [95]. They suggested that there is an enhancement of $T_c$ in $La_{1.48}Nd_{0.4}Sr_{0.12}CuO_4$ near the interface with $La_{1.84}Sr_{0.16}CuO_4$ due to the higher phase stiffness in the later material. Alternatively, they also suggested the possibility of stripe order in $La_{1.84}Sr_{0.16}CuO_4$ induced by the $La_{1.48}Nd_{0.4}Sr_{0.12}CuO_4$ layer.

Altogether, one can conclude that while there is some experimental evidence of interface superconductivity enhancement in $La_{2-x}Sr_xCuO_4$–based heterostructures, at this point the question of underlying mechanism(s) is still largely unsettled.

## 4. Surface superconductivity

### 4.1 Possible surface high-$T_c$ superconductivity in Na-doped $WO_3$

The 5d-transition-metal oxides $WO_3$ and $Na_xWO_3$ are similar in electronic structure to 3d oxides. The crystal structure is similar to that of $ABO_3$ perovskites, with $W^{6-}$ ions occupying the octahedral B cation sites. The A cation sites in $WO_3$ are vacant, while in $Na_xWO_3$ they are occupied by $Na^+$ ions. Stoichiometric $WO_3$ is an insulator, since the W 5d band is empty; when $Na^+$ ions are added to $WO_3$, they donate their 3s electrons to the W 5d band, resulting in bulk metallic behavior for $x < 0.3$ [96]. These materials in tetragonal or hexagonal form exhibit bulk superconductivity at sub-liquid helium temperature [97-101].

However, in 1999 S. Reich and Y. Tsabba claimed a discovery of superconductivity with $T_c \approx 90$ K in $WO_3$ heavily doped with $Na^+$ in a few-nanometers-thick layer near the surface. Pristine $WO_3$ samples did not show any antiferromagnetic background, while the doped samples showed a diamagnetic signal (see figure 20) which the authors attributed to 2D superconducting islands at the surface of the sample. The group backed this up by scanning tunneling spectroscopy (STS) measurements [102,103]. The samples were covered by a layer of gold, thick enough to provide continuous conductive coverage, but thin enough for the DOS at the Au surface to be influenced by superconducting regions below, *via* a proximity effect. About 90% of the material was found to be insulating while the remaining 10% consisted of localized islands exhibiting a gap $\Delta \approx 16$ meV in DOS, with a variation of 30% among different islands. The size of the regions with SC-like gap structures ranged from 20 nm to 150 nm; this non-uniformity was attributed to inhomogeneity in Na doping. Regrettably, so far no other group was able to reproduce the direct observation of HTS in Na:$WO_3$, and hence this result remains questionable.

### 4.2. Field effect

The free charge carrier density is a key parameter of the electronic state of condensed matter, and the ability to change it quasi-continuously is critical in *e.g.* study of quantum phase transitions. The most common method used to change the carrier density of a crystal is by chemical substitution or interstitial doping. However it is not easy to change and measure the doping level with enough precision; in addition, this process may not be reversible and may induce structural modifications and disorder that complicate comparison between samples.

The electric field effect has been gaining popularity as an alternative method to perform such experiments, allowing controlled, reversible and virtually continuous changes of the carrier concentration without adding disorder. However, this method has serious limitations of its own: the characteristic thickness of the accumulation or depletion layer is determined by the electrostatic screening length, which in the semi-classical metallic limit is the Thomas–Fermi length ($\lambda_{TF}$), and which is extremely short, on the order of 1 Å, in common metals. Hence one needs extremely thin – just few atoms thick - yet high quality films, which are rarely feasible and never easy to make. In addition, for a significant variation in the surface charge density one needs an extremely large electric field. This imposes harsh requirements on the gate insulator; *e.g.*, $SiO_2$, the dielectric that is currently used in Si-based field effect transistors (FETs), at best allows a change of $2\times10^{13}$ carriers/cm$^2$ [104].

The first experiments with the electric field effect on superconductivity were performed over 50 years ago, when Glover and Sherrill [105] induced a change of the order of $10^{-4}$ K in $T_c$ of indium and tin, using a parallel plate capacitor fabricated with mica. In the last decade, the tuning range has greatly increased thanks to advances in techniques of deposition of oxides with high dielectric constants such as $HfO_x$ and $SrTiO_3$, as well as ferroelectrics such as $Pb(Zr,Ti)O_3$, and most recently, by using electrochemical cells in which a solid gate insulator is replaced by a polymer / ionic salt electrolyte or by an ionic liquid. (Figure 21). When a gate voltage is applied, the mobile ions of opposite signs contained within the electrolyte accumulate at the surface of the film and on the gate electrode. Such devices – dubbed electrolyte double layer transistors (EDLTs) - can sustain, within the Helmholtz double layers, electric fields in excess of $10^8$ V/cm, with the induced surface charge density reaching $10^{14}$ - $10^{15}$ cm$^{-2}$. In what follows we will briefly review FET and EDLT experiments on several low-$T_c$ and high-$T_c$ superconductors.

***Theoretical studies.*** A great deal of theoretical work has been carried out in order to describe the superconducting field effect transistors (SuFET). Brazovskii and Yakovenko concluded already in 1988 that it should be possible to induce HTS on the surface of insulating $La_2CuO_4$ and $YBa_2Cu_3O_{7-\delta}$ [106,107]. More recently, much work has been carried out by Pavlenko *et al*. focusing first on ferroelectric/superconductor structures containing $Pb(Zr,Ti)O_3$ or $Ba_xSr_{1-x}TiO_3$ and $YBa_2Cu_3O_{7-\delta}$ [108-111]. The study of ferroelectric-superconductor interaction led to the phase diagrams of the system as a function of the coupling between the carriers and the polarization of the ferroelectric layer. It was shown that an important interaction exists between superconductivity and ferroelectricity leading to, for example, suppression of superconductivity or appearance of ferroelectric domains. A trilayer SuFET device was also proposed in which the superconducting layer is sandwiched between two ferroelectric layers. This configuration should allow increased carrier accumulation in one of the interfaces enabling a larger field-induced modulation of the critical temperature. As the gate material, $SrTiO_3$ was proposed, in order to avoid spontaneous polarization of ferroelectrics, which leads to a reduction of the carrier accumulation. Subsequently, Pavlenko with coworkers turned to study of insulator/superconductor heterostructures $SrTiO_3/YBa_2Cu_3O_{7-\delta}$ with the gate voltage applied across STO. [47,112-114]. The different response of $T_c$ in underdoped and overdoped cuprates under an applied electric field in SuFET devices was attributed to orbital reconstruction and phonon coupling at the interface.

***Amorphous bismuth.*** Parendo *et al*. [115] were able to induce an *S-I* transition in amorphous bismuth layers using the field effect with $SrTiO_3$ as the gate dielectric, see figure 22. The films were 10 Å thick and were initially insulating. The applied voltage was tuned between 0 and 42.5 V, increasing the sheet carrier density by $3\times10^{13}$ $cm^{-2}$. The dependence of $T_c$ on the carrier density was ascribed to an increase of screening in Bi, which should reduce the repulsive Coulomb interaction among electrons, and increase the relative importance of the phonon-mediated attractive interaction that gives rise to superconductivity.

***$SrTiO_3$.*** Compared to HTS cuprates and conventional superconductors, the great advantage of $SrTiO_3$ is that the carrier concentration required to achieve superconductivity is 2-3 orders of magnitude smaller ($\approx10^{18}$ $cm^{-3}$), and for this reason it is considered an ideal candidate for field-effect experiments. Takahashi *et al*. [116] studied the *S-I* transition in Nb doped $SrTiO_3$. Epitaxial heterostructures composed by 50 nm thick layer of ferroelectric $Pb(Zr_{0.2}Ti_{0.8})O_3$ (PZT) and a 26 nm thick layer of superconducting $Sr(Ti_{0.98}Nb_{0.02})O_3$ were grown on (001) $SrTiO_3$ single crystal substrates. The polarization of PZT was switched locally by using an AFM tip and applying $V_g = \pm12$ V. The result is summarized in figure 23: at about 270 mK, switching of the ferroelectric polarization induces a transition from the normal state ($P^+$) to a zero resistance superconducting state ($P^-$). The inset of figure 23 shows the piezo-response image of a part of the conducting path after switching the polarization to the $P^+$ state.

Caviglia *et al*. studied LAO/STO heterostructure with contacts in the 'transistor' configuration, modulating the carrier density by applying $V_g = \pm$ 300 V across the STO dielectric [117]. The rich phase diagram of this system is illustrated in figure 24. A 2D superconducting dome reminiscent of that found in HTS cuprates is separated from an insulating phase by a quantum critical point (QCP). Close to this QCP, the transition line $T_{BKT}$ which separates the superconducting and metallic phases exhibits quantum critical scaling $T_{BKT} \sim (\delta V_g)^{z\nu}$, where $V_g$ is roughly proportional to the carrier density, and the product of critical exponents is $z\nu = 2/3$. A similar exponent product was found in 2D *S-I* quantum phase transitions in amorphous bismuth [115] and in $Nb_{0.15}Si_{0.85}$ films [118], and is indicative of so-called 3D XY scaling. Furthermore, the observation of significant negative magnetoresistance in the insulating phase hinted at the presence of weak localization in this region of phase space.

In 2008 Ueno *et al*. induced superconductivity in a $SrTiO_3$ single crystal using polyethylene oxide containing $KClO_4$ as the gate electrolyte [119]. This pioneering work opened the path to subsequent application of the same technique to a number of other materials, some of which are listed below.

Another important advance was reported quite recently by3 Park *et al*., who were able to grow $LaAlO_3/SrTiO_3$ nanowires with atomically flat interfaces on silicon [120]. In principle, this opens the door to integration of correlated-electron effects with Si electronics. [121].

***$YBa_2Cu_3O_{7-x}$ and related compounds.*** Starting two decades ago, much work has been done on field-effect devices using this prototypical HTS compound and the proven FET technology [122-126]. This work will not be reviewed here since an excellent and comprehensive review is available [124]. We will mention here only some recent work using new ideas and techniques.

Ahn *et al*. [127] used off-axis radio-frequency magnetron sputtering and grew epitaxially on a (001)-oriented single-crystal STO substrate first a 72 Å thick $PrBa_2Cu_3O_7$ buffer layer, followed by a 1-2 UC (12-24 Å) thick layer of heavily underdoped $GdBa_2Cu_3O_{7-x}$, and capped with a 3,000 Å thick layer of a ferroelectric oxide, $Pb(Zr_xTi_{1-x})O_3$ (PZT). By switching the polarization field of PZT, they were able to modulate the doping level in $GdBa_2Cu_3O_{7-x}$ and change $T_c$ by 7 K, which enabled them to demonstrate the *S-I* transition

Dhoot *et al*. used ionic liquid gating [128] and were able to tune the critical temperature of $YBa_2Cu_3O_{7-x}$ by more than 10 K using 1-ethyl-3-methylimidazolium bis(trifluoromethylsulfonyl) imide (EMIM:TFSI) and poly(ethylene oxide)/lithium perchlorate ($PEO/LiClO_4$). However, they reported a lack of reversibility and reproducibility, probably attributable to chemical reaction between the electrolyte and the film.

***ZrNCl.*** This material is a semiconductor with an optical bandgap of about 3 eV, and it has a layered structure consisting of ZrN layers sandwiched between two chlorine layers. Upon intercalation of Li, electrons are transferred from Li atoms to the ZrN layers. $Li_xZrNCl$ is a superconductor with $T_c$ between 11 and 15 K for $0.06 < x < 0.37$ [129,130]; this is 2-6 K higher than the $T_c$ of ZrN although in both cases superconductivity occurs in the ZrN layers. Ye *et al*. applied the EDLT technique to ZrNCl using N,N-diethyl-N-(2-methoxyethyl)-N-methylammonium bis (trifluoromethyl suphonyl)imide (DEME-TFSI) as electrolyte (Figure 25) and achieved a modulation of the carrier density of up to $2.5\times10^{14}$ cm$^{-2}$. Superconductivity was shown to be confined to two topmost layers of ZrN (aprox. 20 Å in thickness) and subsequently confirmed by magnetic measurements [131].

***$La_{2-x}Sr_xCuO_4$.*** Atomic-layer-by-layer MBE described in Section 2.3 was used to synthesize a number of 1, 1.5 or 2 UC thick films of $La_{2-x}Sr_xCuO_4$ with x = 0.06 to 0.20, over some insulating $La_2CuO_4$ buffer layers, on single-crystal $LaSrAlO_4$ substrates [132]. Using either polymer electrolytes or ionic liquids, EDLT devices of well-defined geometry were fabricated lithographically for accurate measurements of resistivity and magnetic susceptibility. Upon applying a gate voltage, large shifts were observed both in the normal state resistivity and in $T_c$, see figure 26. In the top layer, shifts were seen in the induced carrier density $x$ per Cu by up to ±0.04 and shifts in $T_c$ (defined as the transition midpoint) by as much as 30 K – more than 75% of the full range from about 2 K to the maximum ($T_c^{max} \approx 40$ K) observed in thicker single-phase LSCO films at optimum doping. The field effect works in both directions; it is possible to achieve electron accumulation or depletion, and increase or decrease in $T_c$, in the expected direction, depending on whether the sample is initially overdoped or underdoped. Diamagnetic response, measured by two-coil mutual inductance technique, also showed large shifts in $T_c$ when the gate voltage was applied, see figure 26b. This demonstrates complete superconductivity - strong diamagnetic screening onsets when the supercurrent closes a complete loop at or below $T_c(R=0)$.

With proper choice of electrolytes, charging time and temperature, etc., the process was found to be reversible and non-destructive, unless some threshold voltage is exceeded, so it was possible to record hundreds of resistance vs. temperature and carrier density curves on the same device. These were shown to collapse onto a single function as predicted for a 2D quantum *S-I* transition. The observed critical values of $x$ and the sheet resistance $R_\square$ were $x_c \approx 0.06$ and $R_c = 6.45 \pm 0.10$ kΩ. The later is precisely equal to the quantum resistance for pairs, $R_Q = h/(2e)^2 = 6.45$ kΩ, suggestive of a phase transition driven by quantum phase fluctuations, and Cooper pair (de)localization. The data collapse was excellent up to about 10 K for critical exponent product $z\nu = 1.5 \pm 0.1$, and the exponent was found to be identical on both sides of the S-I transition.

***Open questions.*** How ionic liquid interacts with the material's surface remains a question. In most studies published so far the authors simply assume that this interaction is strictly electrostatic, and that there is neither incorporation of ions nor generation of vacancies in the film as the result of chemistry or when voltage is applied. These assumptions may not always be justified. Accurate compositional analysis, with the depth resolution on an atomic scale, is quite difficult given that the thickness of the channel is just few nanometers, and it is buried under the electrolyte. Even if the electrolyte is removed, it may leave an adsorbed monolayer, and it will be hard to tell whether this is on the surface or in the channel. However, given that this is probably going to remain one of most important questions in EDLT physics, one can expect that in the coming years new techniques will be developed, or the old ones modified and improved, to address this question systematically and quantitatively. In the meantime, one has to rely on circumstantial evidence, such as the fact that the results are reversible, independent on the choice of the ionic liquid, etc.

A more profound question is whether the 2D superconducting phenomena, induced at a surface by electric field or at an interface by charge carrier accumulation and depletion, are the same in all aspects as in the bulk crystals or thicker films of the same material doped to the same level – even if $T_c$ is similar. One can indeed conceive of a number of reasons why they may not be identical. In both cases, interface or surface superconducting layer is exposed to a large local perpendicular electric field. This breaks the inversion symmetry as well as the invariance with respect to a mirror reflection in the interfacial plane. Under these circumstances, Rashba spin-orbit coupling may arise and cause mixing of singlet and triplet superconductivity [133]. More mundane, it can cause substantial local ionic displacement, making the local crystallographic structure very different from that of the bulk crystals of the same compound – in turn modifying the electron spectrum and DOS at the Fermi level, the phonon spectrum, the electron-phonon interaction, etc. The landscape of disorder can be also very different; on one side, one would expect much less local disorder here than what would be induced by chemical doping, while on the other side morphological defect such as terracing, island growth and concomitant surface and interface roughness, antiphase grain boundaries, misfit dislocations, etc., could play a much bigger role. Last but not least, oxygen being volatile, there is a possibility that local oxygen landscape is different, with a greatly increased density of either oxygen vacancies or of excess interstitial oxygen, since modified local electrostatics can change the energy of the corresponding oxygen sites.

These and other basic questions about 2D superconductivity at surfaces and interfaces will be very likely a matter of intense research and debate in the forthcoming years.

## 5. Summary and outlook

The quest for interface and surface superconductivity started almost fifty years ago with theoretical proposals by De Gennes [1] and Ginzburg [3] and was followed by relatively slow progress in both experiment and theory work spread over four decades. In the last decade, however, major strides have been made in the technique of atomic-layer engineering - the capability to deposit ultrathin perfect layers with of various materials including complex oxides and high-$T_c$ cuprate superconductors, with atomically sharp interfaces. Controlled chemical modification of even a single, pre-selected atomic layer ($\delta$-doping) has been demonstrated.

These technical advances have enabled breakthrough experiments, performed largely in the last few years, in which interface superconductivity was discovered at a junction of two materials, neither of which was superconducting otherwise. The two prime examples are STO/LAO heterostructures made of two different insulators, and LSCO/LCO heterostructures made of one insulator and one (non-superconducting) metal.

In parallel, quasi-continuous tuning of charge carrier density at the film surface by electric field has been demonstrated in STO, and quite recently, using a new electrolyte-based technique, in cuprates as well. This has enabled some basic physics experiments, *e.g.* establishing the nature of *S-I* phase transitions. Evidence was obtained for quantum ($T$ = 0) phase transitions driven by quantum phase fluctuations. Without doubt, such experiments will be extended in near future and the hope is that we will learn a lot more about 2D superconductivity.

In the above examples, charge accumulation and depletion by interfacial effects or external field is used as a tuning parameter to bring a thin interfacial or surface layer (of $SrTiO_3$ or $La_2CuO_4$) to its optimum doping level, reaching the $T_c^{max}$ that can be obtained in chemically doped bulk samples. However, there are at least two documented cases where an interfacial superconducting layer has $T_c$ larger than $T_c^{max}$, the situation we refer to as *interface-enhanced superconductivity*. One of these are SLs of chalcogenide semiconductors (PbTe/SnTe, PbSe/PbS, etc.), known mostly from the work of N. Fogel and her collaborators, where $T_c$ = 6 K has been reached, while in single-phase compounds $T_c^{max}$ < 1.5 K. The phenomenon has been convincingly related to the existence of networks of misfit dislocations, but little is known yet about the mechanism of enhancement. The other case are *M-S'* or *S-S'* cuprate heterostructures, where $M$ = $La_{1.55}Sr_{0.45}CuO_4$, $S$ = $La_{1.85}Sr_{0.15}CuO_4$, and $S'$ = $La_2CuO_{4+d}$, with $T_c$ = 50-52 K > $T_c^{max}$ ≈ 40 K. These heterostructures were studied in detail by I. Bozovic and coworkers, and a clear structural component (elongation of the $c_0$ lattice constant and of the Cu-apical oxygen bond) was identified. However, again much remains unknown about the enhancement mechanism at the time of this writing.

For deeper understanding of these phenomena, one needs detailed information on the atomistic structure, electronic and phonon states in the buried interface. For these, new techniques of characterization of buried interfaces with atomic resolution are necessary and are being developed. Particularly necessary, but particularly difficult to obtain, is information on the density and vertical profile of oxygen vacancies and interstitial oxygen, in the interfacial region. In bulk, useful information on oxygen vacancies can be obtained expediently from Raman spectroscopy [134], even with lateral resolution on ~1 μm scale [135,136], but improving the vertical ($z$-axis) resolution to the necessary level appears like a daunting challenge.

Ultimately, we may not expect too much in terms of real-life applications of superconductivity in STO with $T_c$ = 0.3 K, nor even in LSCO with $T_c$ = 50 K. However, by studying in depth and understanding these two model systems, we may learn some new physics and in the process develop new techniques for discovering novel and perhaps superior superconductors. The search for materials with a higher $T_c$ has led us to consider heterostructures and surface superconductivity as promising avenue with more versatility to engineer the superconducting properties.

As for the choice of materials and mechanisms, a number of theoretical proposals have been put on board, most of which have yet to be tested and proven operative. Already in 1964 and closely following Little's proposal for exciton-mediated electron pairing in organic 1D systems [137], Ginzburg proposed that excitonic superconductivity could occur in superlattices made by alternating thin metallic and dielectric layers [3,138]. This was one of the first suggestions how to design a novel, high-temperature superconductor. In 1989 2D superconductivity had already been shown to be feasible and Ginzburg extended his proposal, pointing to the possibility to manipulate and control the electron interaction energy near an interface, in comparison with that in the bulk, by depositing overlayers of properly chosen partner materials. Indeed, near the interface between two disparate materials one can encounter large changes in crystal structure, electron-ion and electron-electron interactions, *etc*., which can give rise to a wide range of novel phenomena. More recently, S. Kivelson proposed that interfacial $T_c$ enhancement can be achieved by a proximity effect between an underdoped cuprate (where the pairing is strong) and an overdoped cuprate (where the phase stiffness is supposed to be stronger).

Another line of thinking is to use atomic-layer engineering and synthesize heterostructures in which the electronic state in thin layers of oxide of some element other than copper would be modified, by electron depletion/accumulation and other interfacial effects, in such a way to mimic those of $CuO_2$ planes in HTS cuprates. A number of candidate HTS heterostructures have been proposed including $LaAlO_3$/$LaNiO_3$, $SrFeO_3$/$SrTiO_3$, etc. [139]. Some of these have already been explored experimentally as well, to the best of our knowledge without success (defined as a discovery of superconductivity). Two caveats are in place here. First, in most cases such a suggestion is based, explicitly or not, on certain assumptions or models for the mechanism of HTS in cuprates. However, to this day, there is no consensus on the actual mechanism of HTS and hence there is an obvious danger of hinging on a wrong one. Second, one should bear in mind that there are, in fact, many oxides (nickelates, rhutenates, cobaltates, etc.) which are in all key electronic properties quite similar to those of cuprates – but show no superconductivity [140]. Altogether, at this point all these suggestions are just interesting ideas, and their main merit is that they have already stimulated much experimental research; the other proposals yet to come are likely to have the same effect.

What could be major milestone results and discoveries along these lines? For one, we would like to see interface superconductivity between an insulating cuprate (e.g. LCO) and a non-cuprate metal, ideally one that contains no Sr or Ba, thus eliminating the possibility that interface superconductivity occurs as a result of cation interdiffusion and inadvertent chemical doping. Another would be an additional dramatic increase of $T_c$ in cuprates (say observing $T_c$ >

60 K in $La_{1.85}Sr_{0.15}CuO_4$ system, because that would be out of reach of any non-trivial explanation) or in pnictides; the latter is not going to be easy because the materials are very metallic, which implies a short screening length and necessitates perfect surfaces or ultrathin films. Last but not least, we are awaiting a discovery of surface superconductivity, even with a relatively low $T_c$, by say EDLT technique in a material that cannot be made (or at least that has never been made so far) superconducting by chemical doping. Any of these discoveries would certainly provide a great further impetus to this emerging field, the last one probably more than either of the other two. We hope not to wait too long.

**Acknowledgements.** The work at BNL was supported by the U.S. Department of Energy, Basic Energy Sciences, Materials Sciences and Engineering Division. The work was supported by the National Research Foundation, Singapore through Grant NRF-CRP4-2008-04.

**Figure captions:**

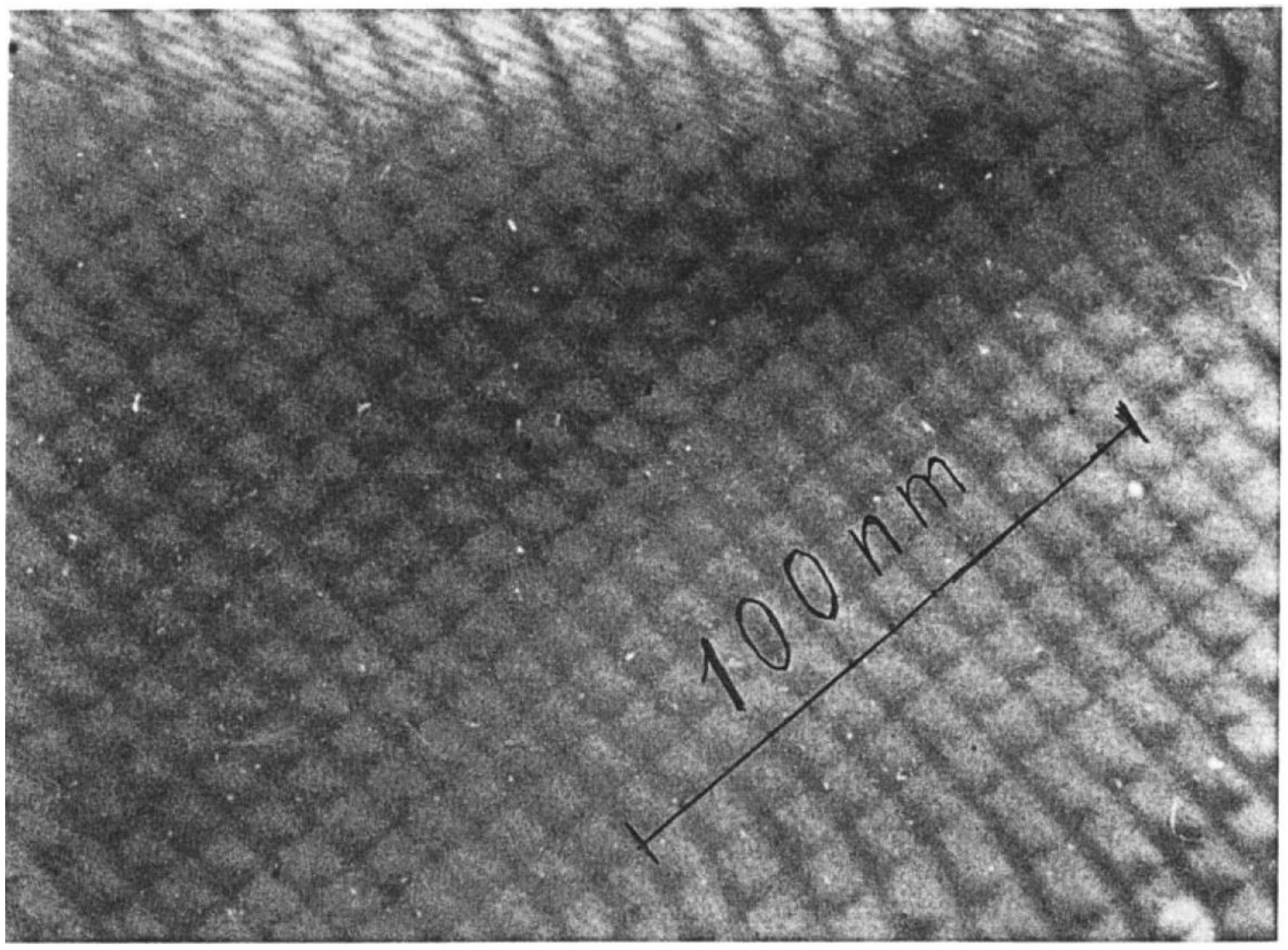

**Figure 1**. Dislocation grid in a PbTe/PbSe bilayer. [141]

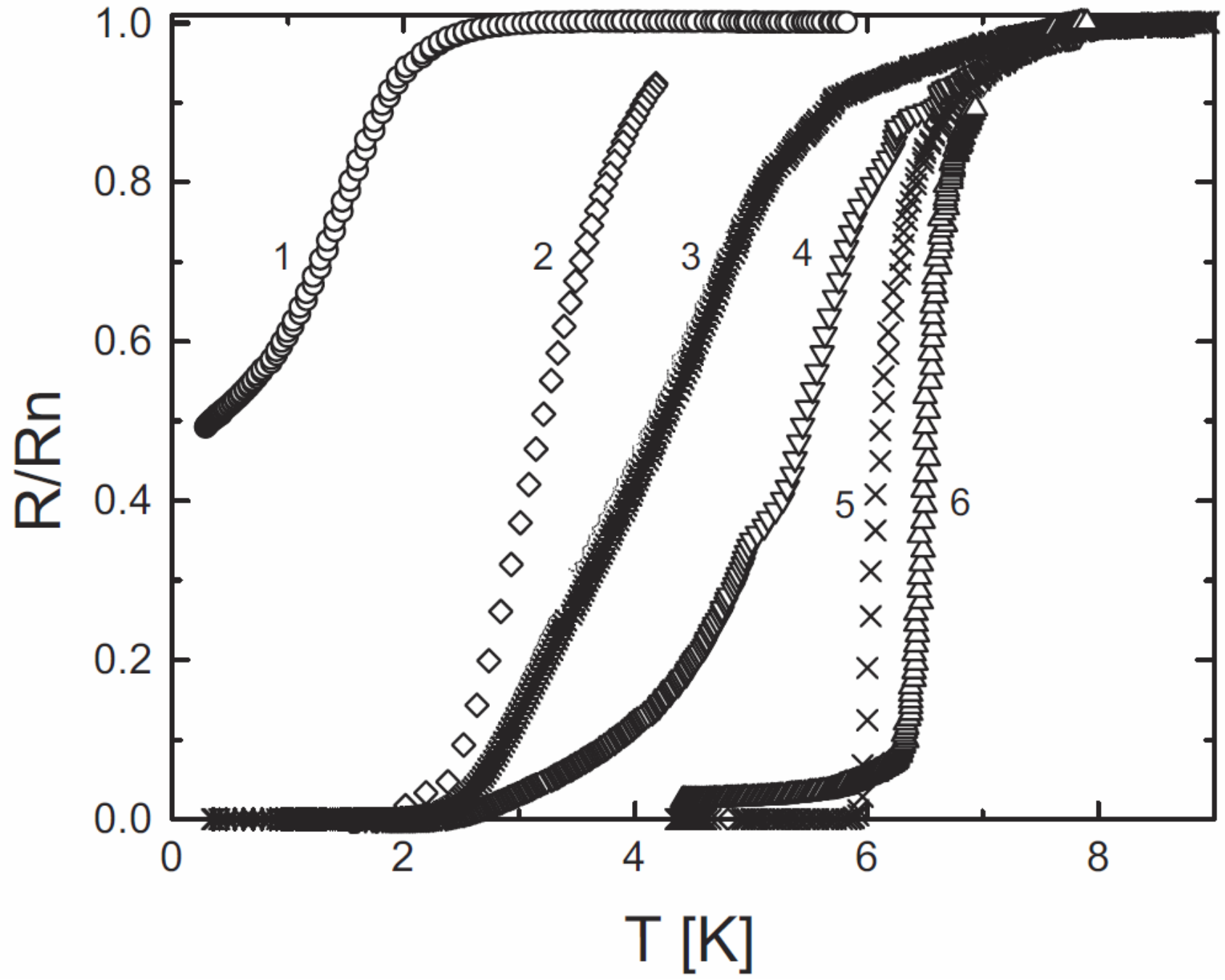

**Figure 2**. Normalized resistance ($R/R_n$) as a function of temperature $T$ for six PbTe/PbS heterostructures. The figure shows the data corresponding to five bilayer heterostructures with different thicknesses: (1) $d_1$ = $d_2$ = 40 nm, (2) $d_1$ =$d_2$ = 100 nm (3,4) $d_1$ = $d_2$ = 80 nm and (6) $d_1$= 200 nm and $d_2$= 40 nm. (5) represents the results for one superlattice with $d_1$ = $d_2$ = 120 nm. [142]

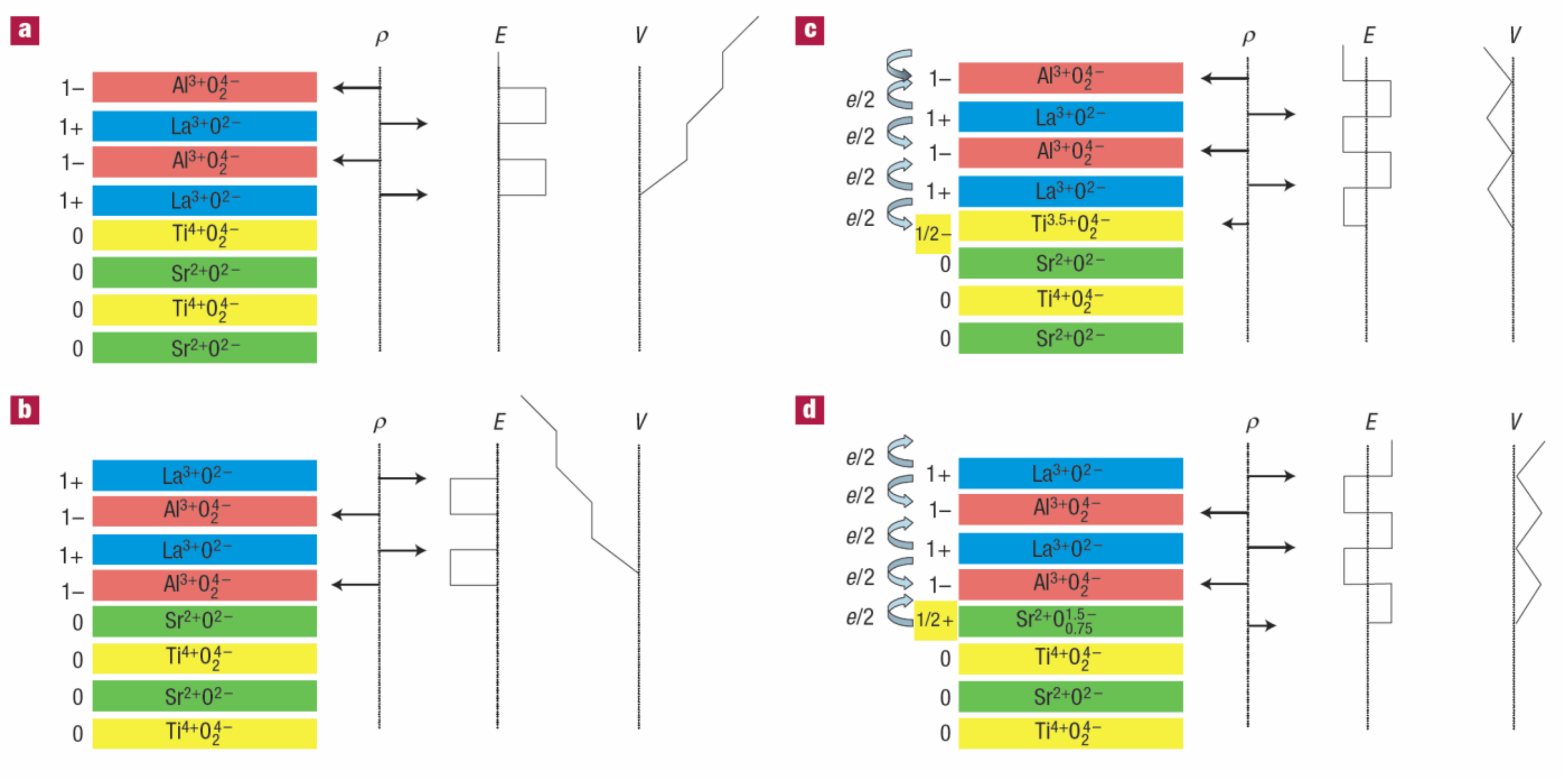

**Figure 3**. The polar catastrophe is illustrated for atomically abrupt interfaces between $LaAlO_3$ and $SrTiO_3$. **a**) The unreconstructed interface has neutral (001) planes in $SrTiO_3$, but the (001) planes in $LaAlO_3$ have alternating net charges (ρ). If the interface layer is $AlO_2$/LaO/$TiO_2$, this produces a non-negative electric field (E), leading in turn to an electric potential (V) that diverges with thickness. **b**) If the interface is instead placed at the $AlO_2$/SrO/$TiO_2$ plane, the potential diverges negatively. **c**) The divergence catastrophe at the $AlO_2$/LaO/$TiO_2$ interface can be avoided if half an electron is added to the last Ti layer. This produces an interface dipole that causes the electric field to oscillate about 0 and the potential remains finite. **d**) The divergence for the $AlO_2$/SrO/$TiO_2$ interface can also be avoided if oxygen vacancies are created to remove half an electron charge from the SrO plane. [36]

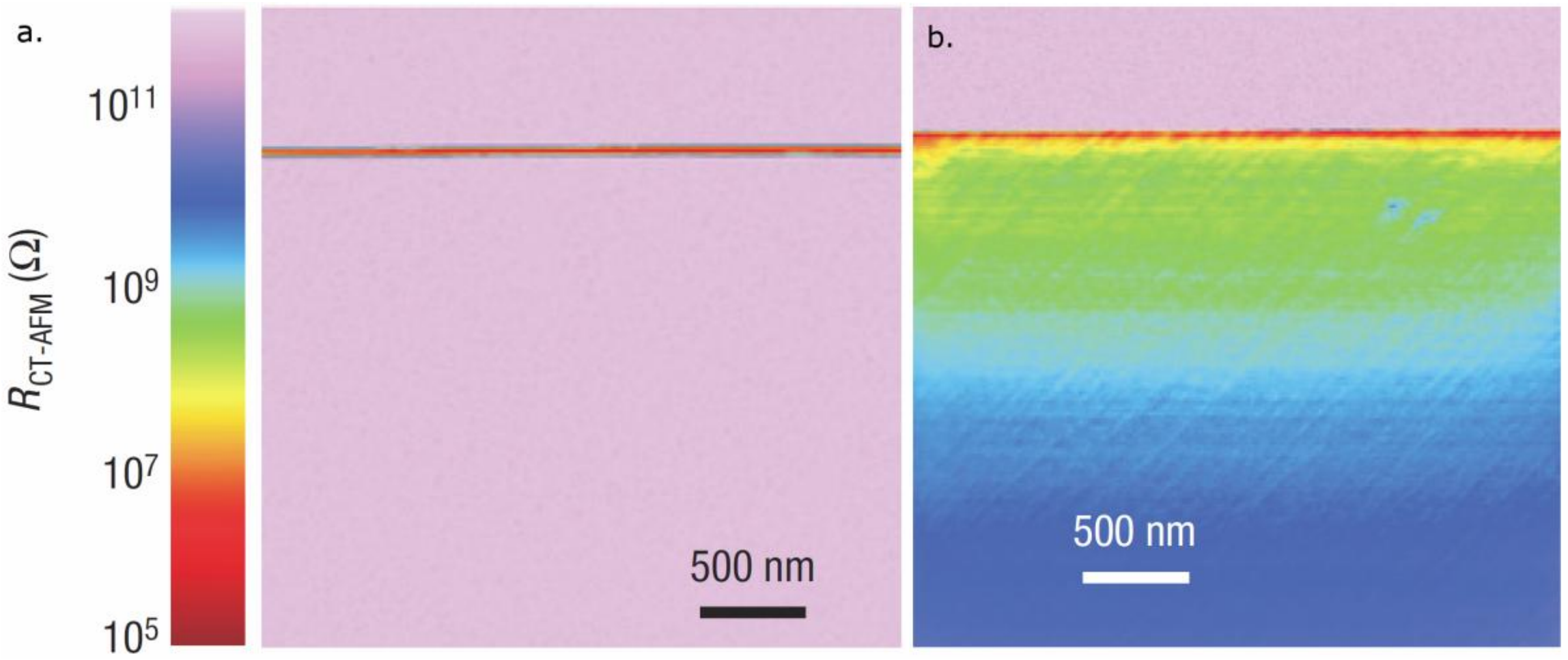

**Figure 4**. Conductivity map around the interface between $LaAlO_3$ and $SrTiO_3$ obtained by Conductive tip Atomic Force Microscopy in cross sectional configuration. **a**) Sample grown at 750 ºC and oxygen pressure $P_{O2} = 10^{-6}$ mbar and cooled down to room temperature at 300 mbar. **b**) Sample grown in the same conditions but cooled down to room temperature at $P_{O2} = 10^{-6}$ mbar. [41]

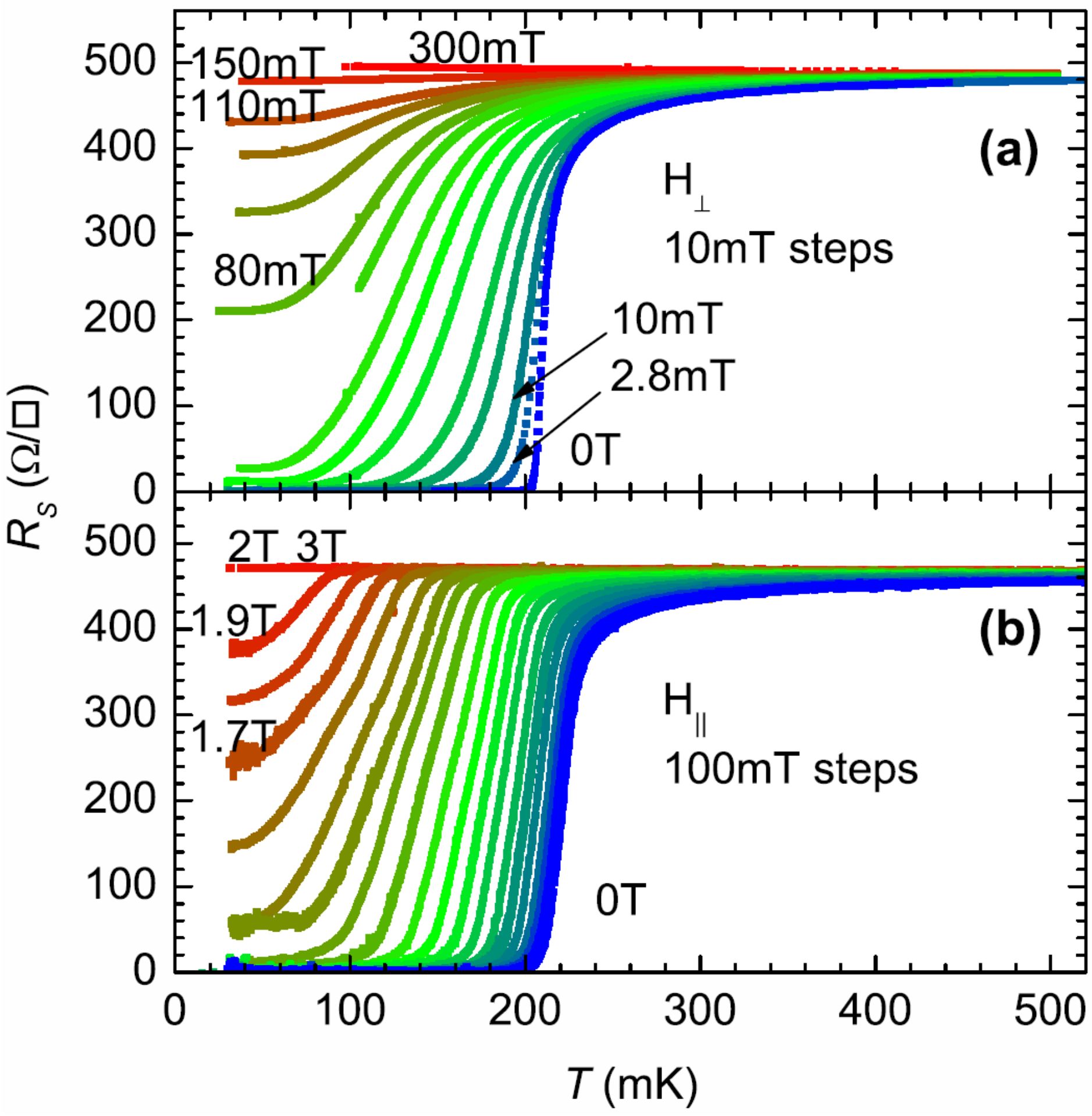

**Figure 5**. Anisotropy of the sheet resistance versus temperature for different magnetic fields in $LaAlO_3$/$SrTiO_3$ structures, **a**) perpendicular and **b**) parallel to the interface. [48]

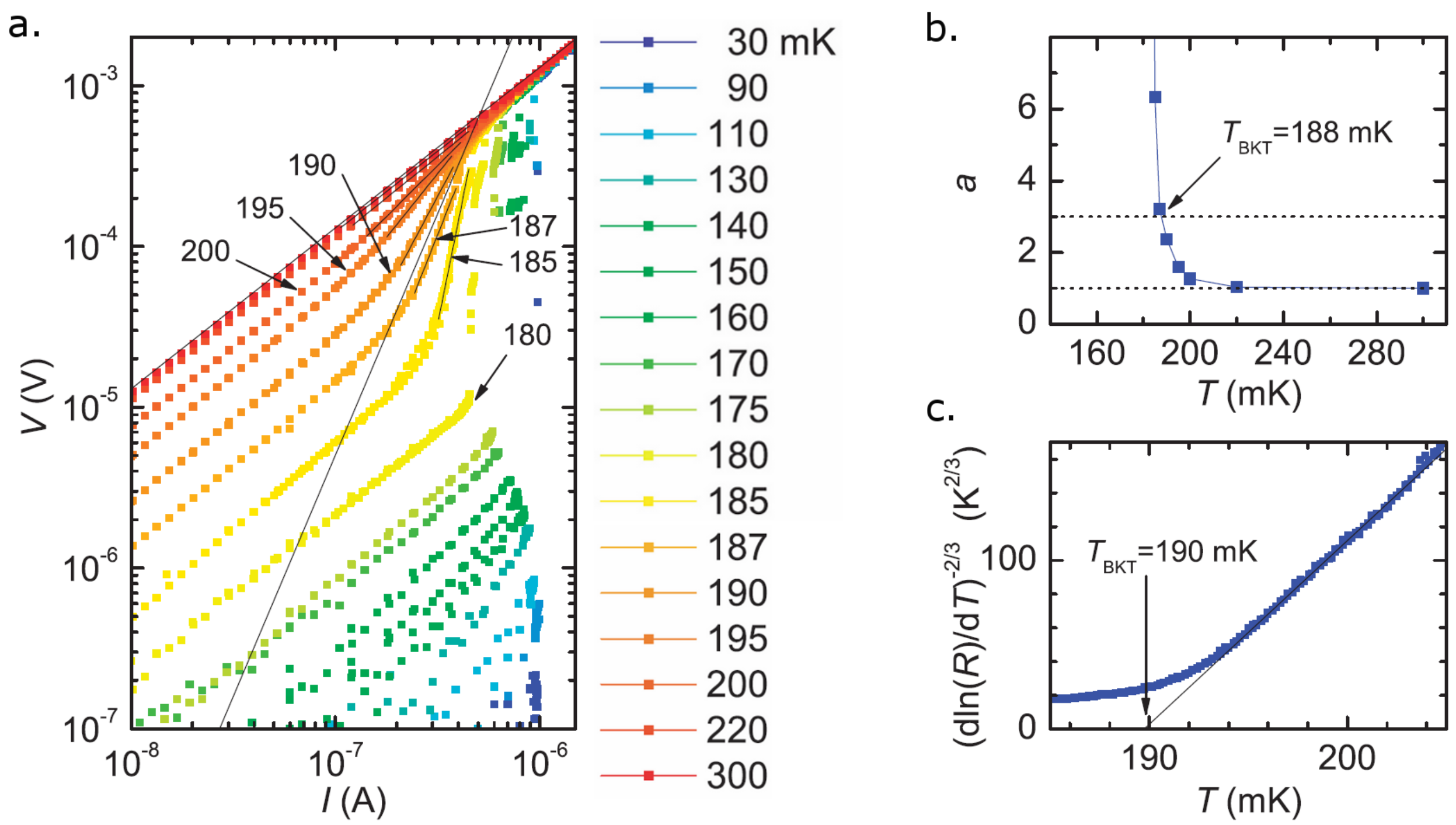

**Figure 6**. Low-temperature transport properties of a heterostructure consisting of 8 UC thick $LaAlO_3$ layer grown on $SrTiO_3$. **a**) Current–voltage curves on a logarithmic scale. The numbers indicate the temperature at which the curves were measured. The two long black lines correspond to $V = RI$ and $V \sim I^3$ dependencies and show that 187 mK < $T_{BKT}$ < 190 mK. **b**) Temperature dependence of the power-law exponent in $V \sim I^a$, as deduced from the fits shown in panel a. **c**) Temperature dependence of the resistance of the same sample for $I$ = 100 nA, plotted on the $[dln(R)/dT]^{-2/3}$ scale. The solid line represents the behavior expected for a BKT transition with $T_{BKT}$ = 190 mK. [11]

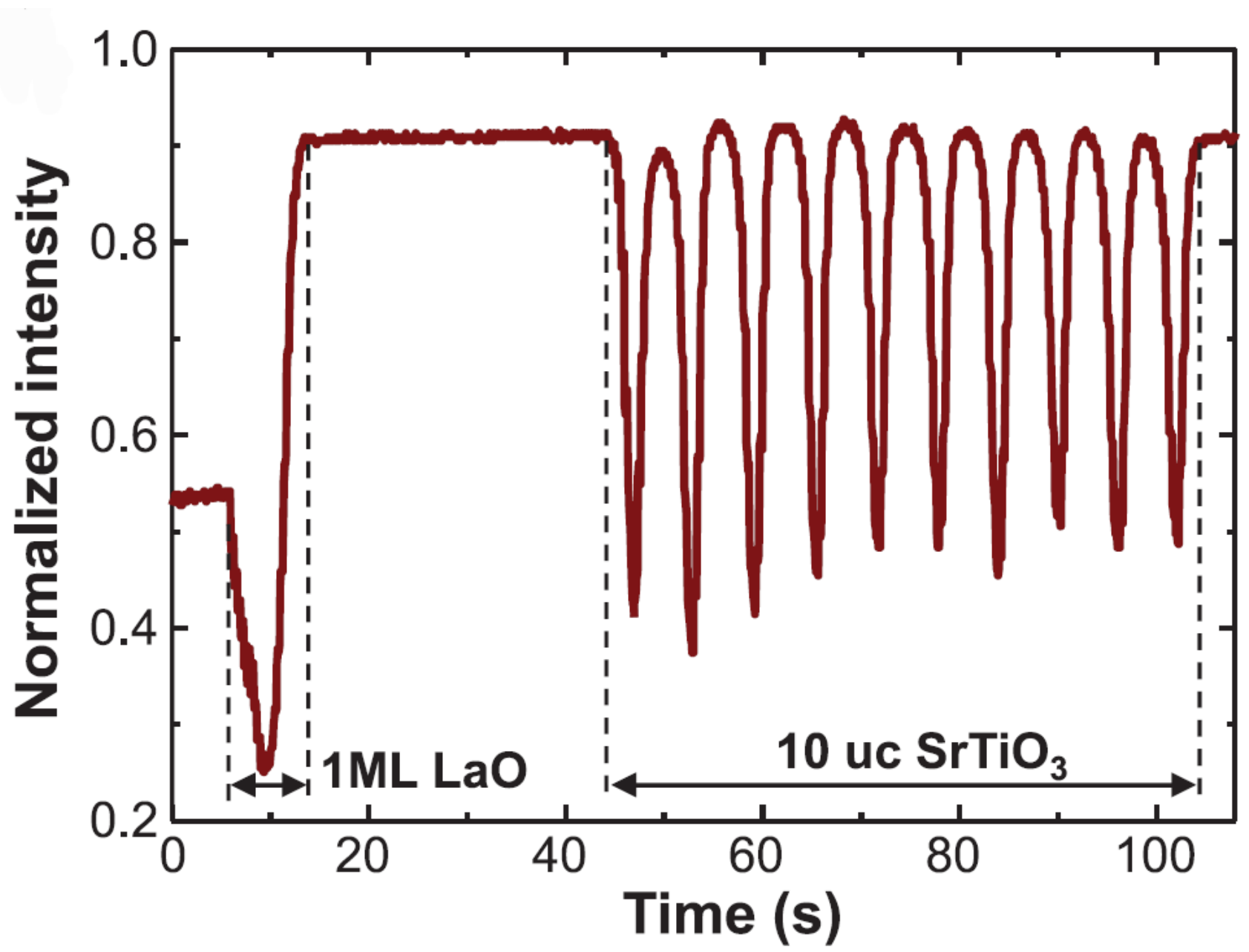

**Figure 7**. Typical RHEED oscillations during growth of one "molecular layer" (1ML) LaO and 10-UC thick $SrTiO_3$ layer on a $TiO_2$-terminated $SrTiO_3$ substrate. [53]

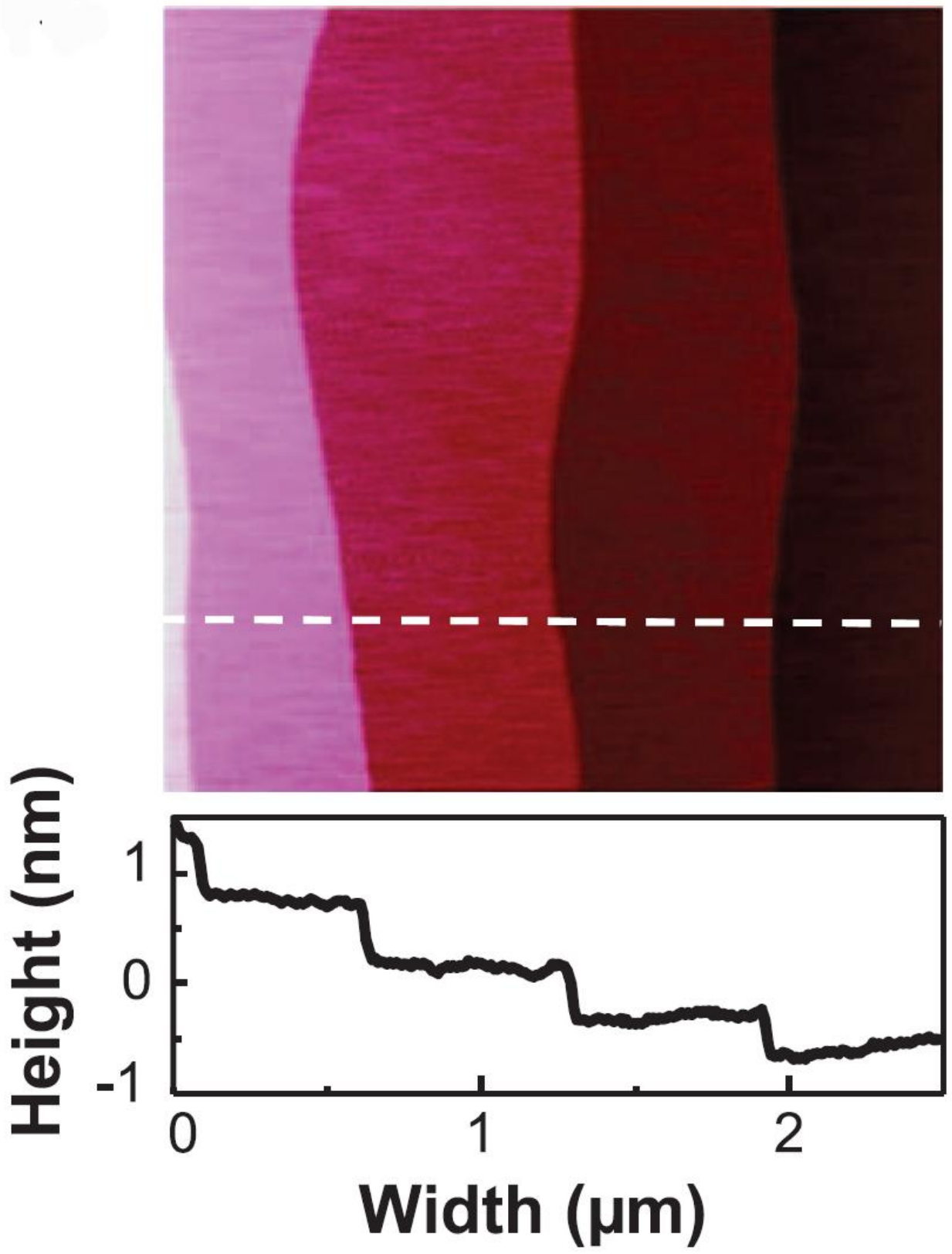

**Figure 8**. AFM image of a 10-UC SrTiO3/1-ML LaO/$SrTiO_3$ heterostructure showing an atomically smooth surface. [53]

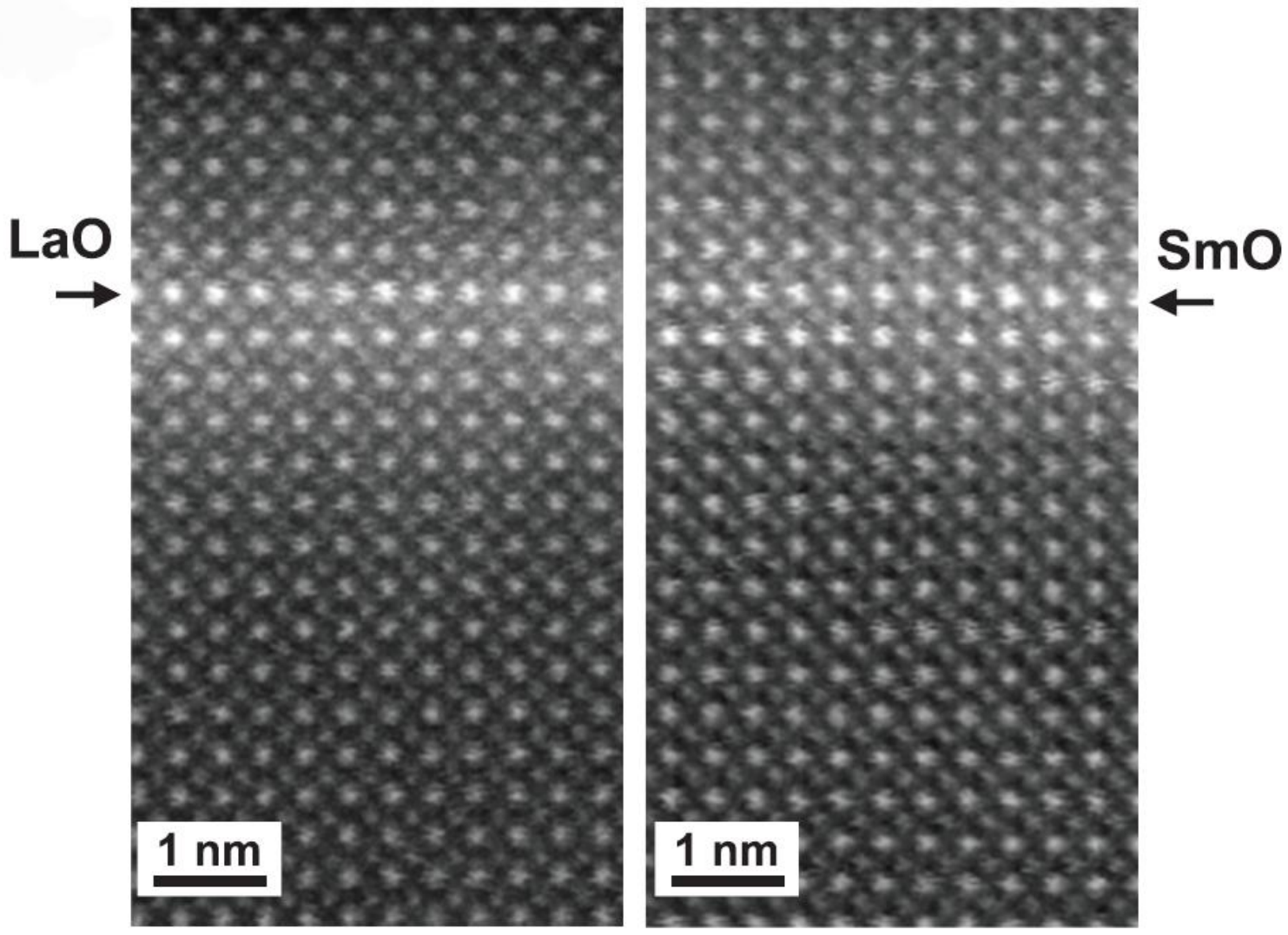

**Figure 9**. High-angle annular dark field images of 10-UC $SrTiO_3$ / 1-ML LaO / $SrTiO_3$ and 10-UC $SrTiO_3$ / 1-ML SmO / $SrTiO_3$ heterostructures. Neither sample shows obvious defects or dislocations, indicating coherent interfaces. [53]

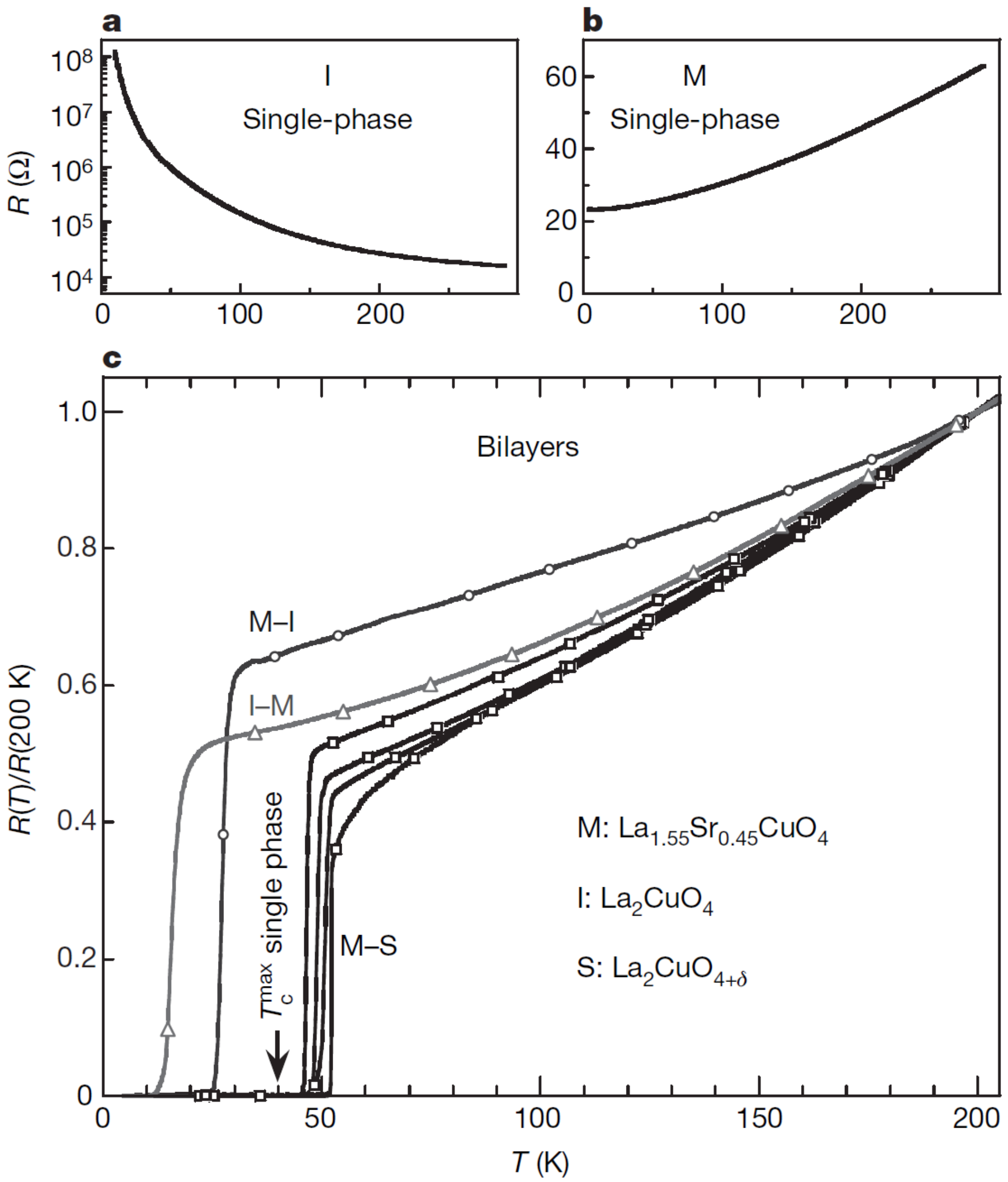

**Figure 10**. Normalized resistance, *R(T)/R(200 K)*, as a function of temperature (*T*) of single-phase $La_{1.55}Sr_{0.45}CuO_4$ (*M*) and $La_2CuO_4$ (*I*) films and $La_{1.55}Sr_{0.45}CuO_4/La_2CuO_{4+\delta}$ (M-*S*), $La_{1.55}Sr_{0.45}CuO_4/La_2CuO_4$ (*M-I*) and $La_2CuO_4/La_{1.55}Sr_{0.45}CuO_4$ (*I-M*) bilayer structures. [60]

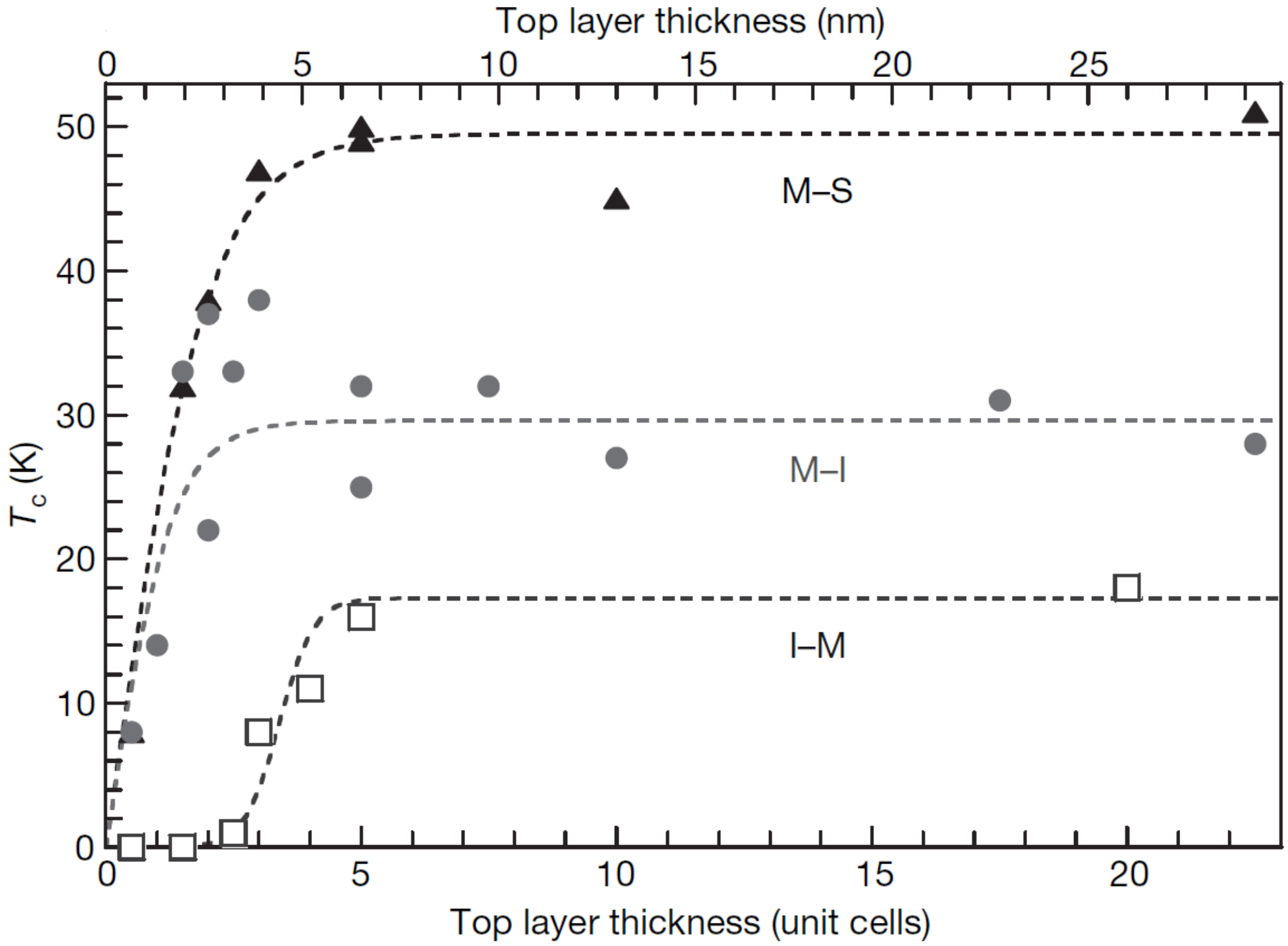

**Figure 11**. Dependence of $T_c$ (defined as the midpoint of the resistive transition) of bilayer structures as a function of the top layer thickness. The bottom layer thickness is fixed at 40 UC (~52 nm). The dashed lines are guides for the eye. [60]

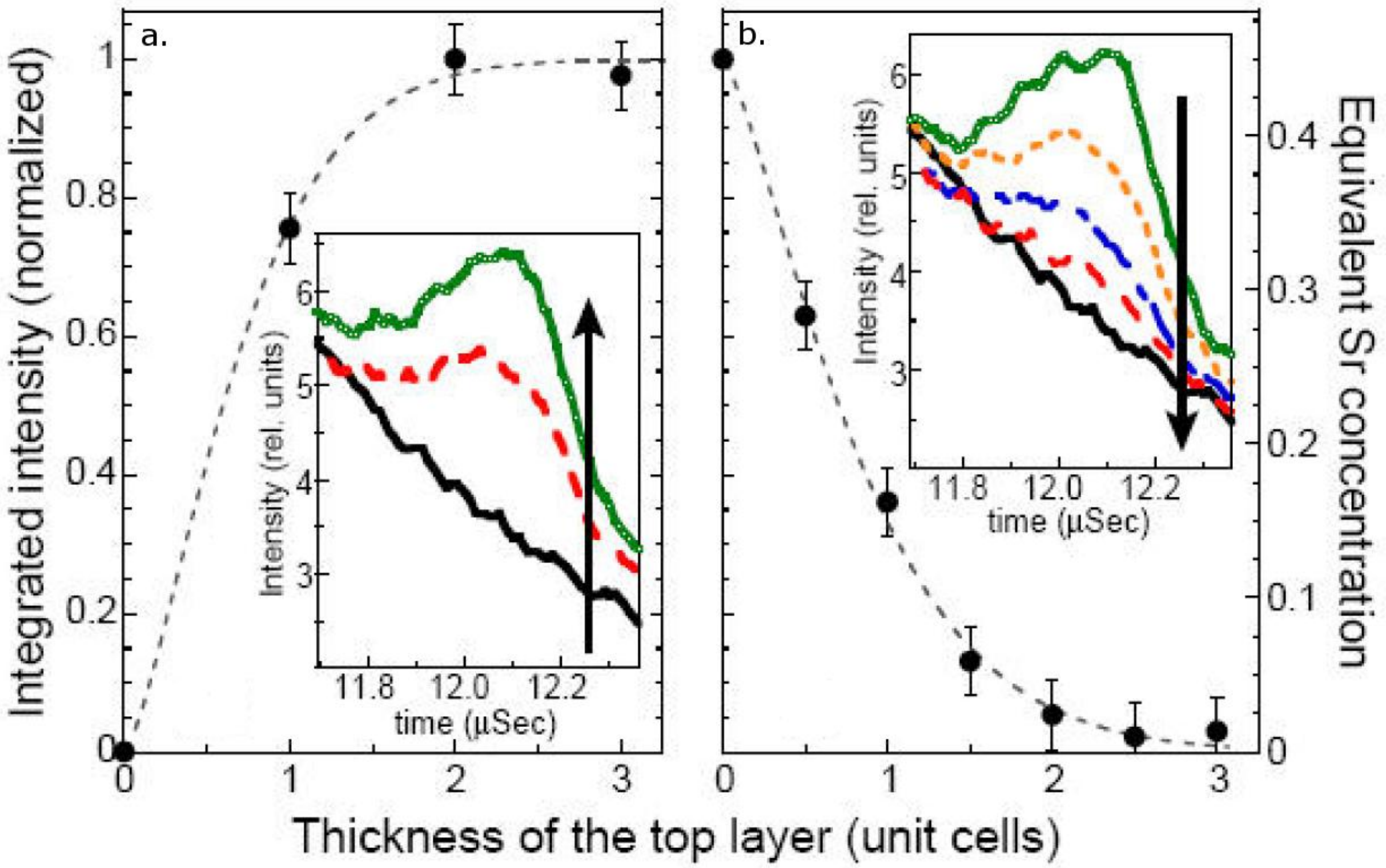

**Figure 12**. Evolution of the normalized integrated intensity of the Sr recoil spectra obtained by Time-of-flight ion scattering and recoil spectra (TOF-ISARS). **a**) The spectra measured during the deposition of $La_{1.55}Sr_{0.45}CuO_4$ on top of $La_2CuO_4$, with the increment of 0.5 UC. The dashed lines are guides for the eye. **b**). The spectra measured during the deposition of $La_2CuO_4$ on top of $La_{1.55}Sr_{0.45}CuO_4$. The Sr recoil peak vanishes after the deposition of 2 UC of the $La_2CuO_4$ top layer. [143]

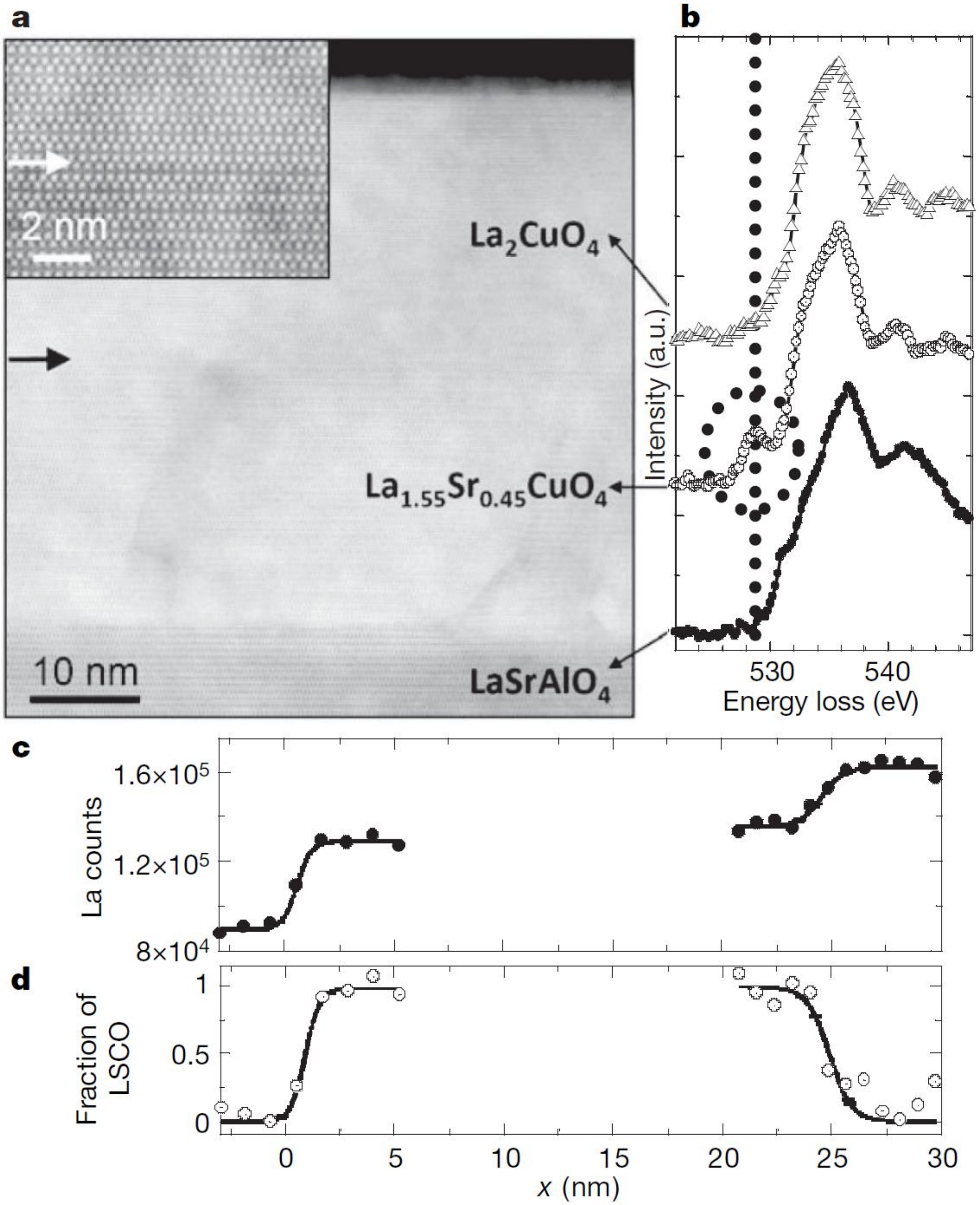

**Figure 13**. Analysis of an *M-I* bilayer by scanning transmission electron microscopy and electron energy-loss spectroscopy. **a**) Annular dark field image of the structure. The inset shows a magnified image of the M-I interface (indicated by an arrow). **b**) Oxygen-K (O-K) electron energy-loss spectra of the three oxides in the structure. An O-K edge pre-peak is observed in the $La_{1.55}Sr_{0.45}CuO_4$ layer (circled). This peak evolves and scales with the doping level (a.u. = arbitrary units). **c**) Integrated La intensity across the bilayer. The La profile increases from the substrate to the *M* layer, and again from the *M* to the *I* layer. **d**) Results of a principal-components analysis of the two interfaces. The fraction of one of two components, corresponding to the O-K edge in $La_{1.55}Sr_{0.45}CuO_4$ is shown. [60]

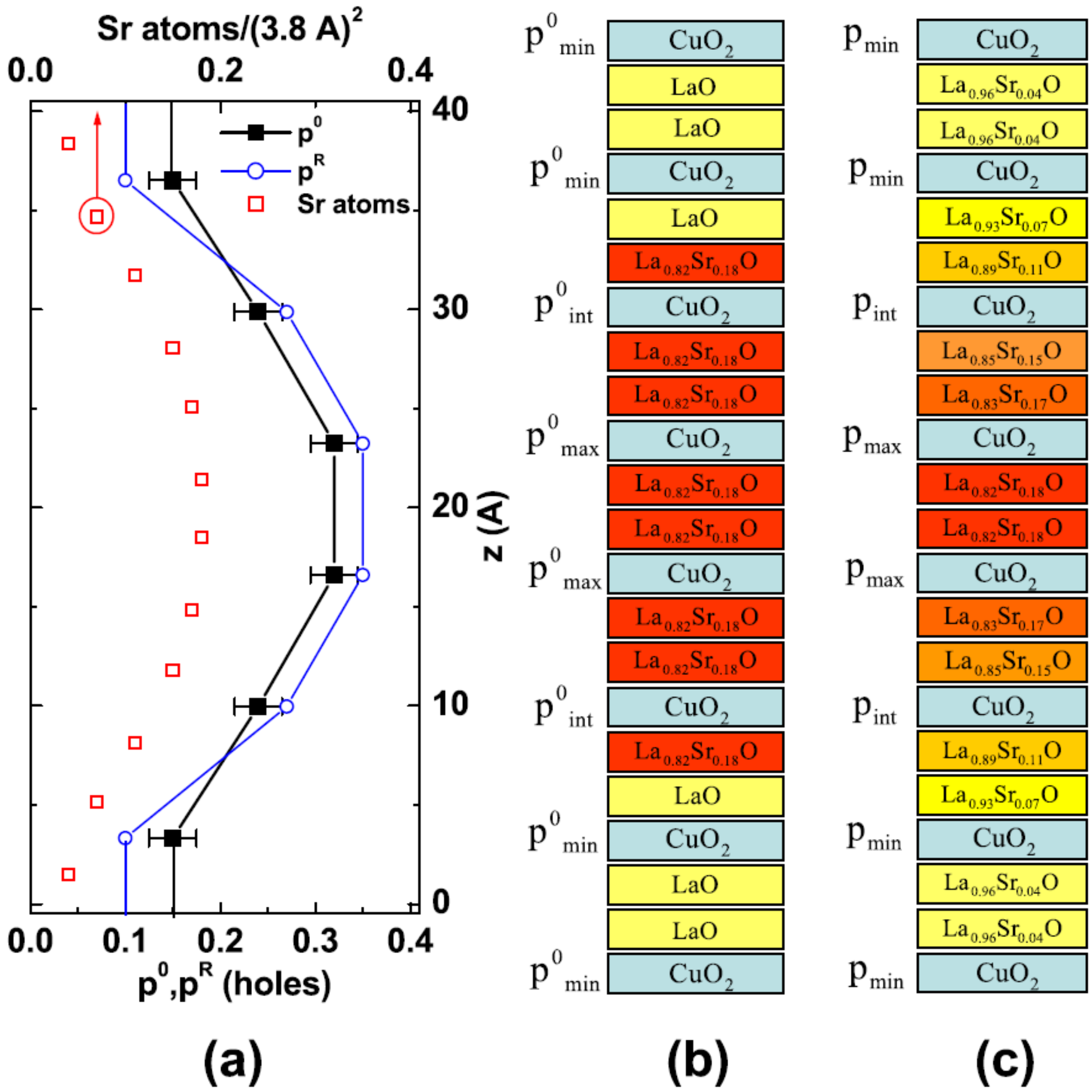

**Figure 14**. **a**) Layer-resolved hole count in an ideal structure with no interface roughness (black solid squares), and the hole count obtained from structural roughness only (blue open circles). These two distributions are convolved to obtain the hole distribution in real structure (panel c). The nominal distribution of $Sr^{2+}$ ions is also represented (red open squares). **b**) Sketch of the superlattice hole distribution in an ideal structure, aligned to "panel a" for comparison. **c**) Sketch of the hole distribution in real structure, accounting for La/Sr interdiffusion. $p^0_{min}$ and $p^R_{min}$ are the hole counts in the LCO layers contributed by electronic effects and by roughness, respectively. [63]

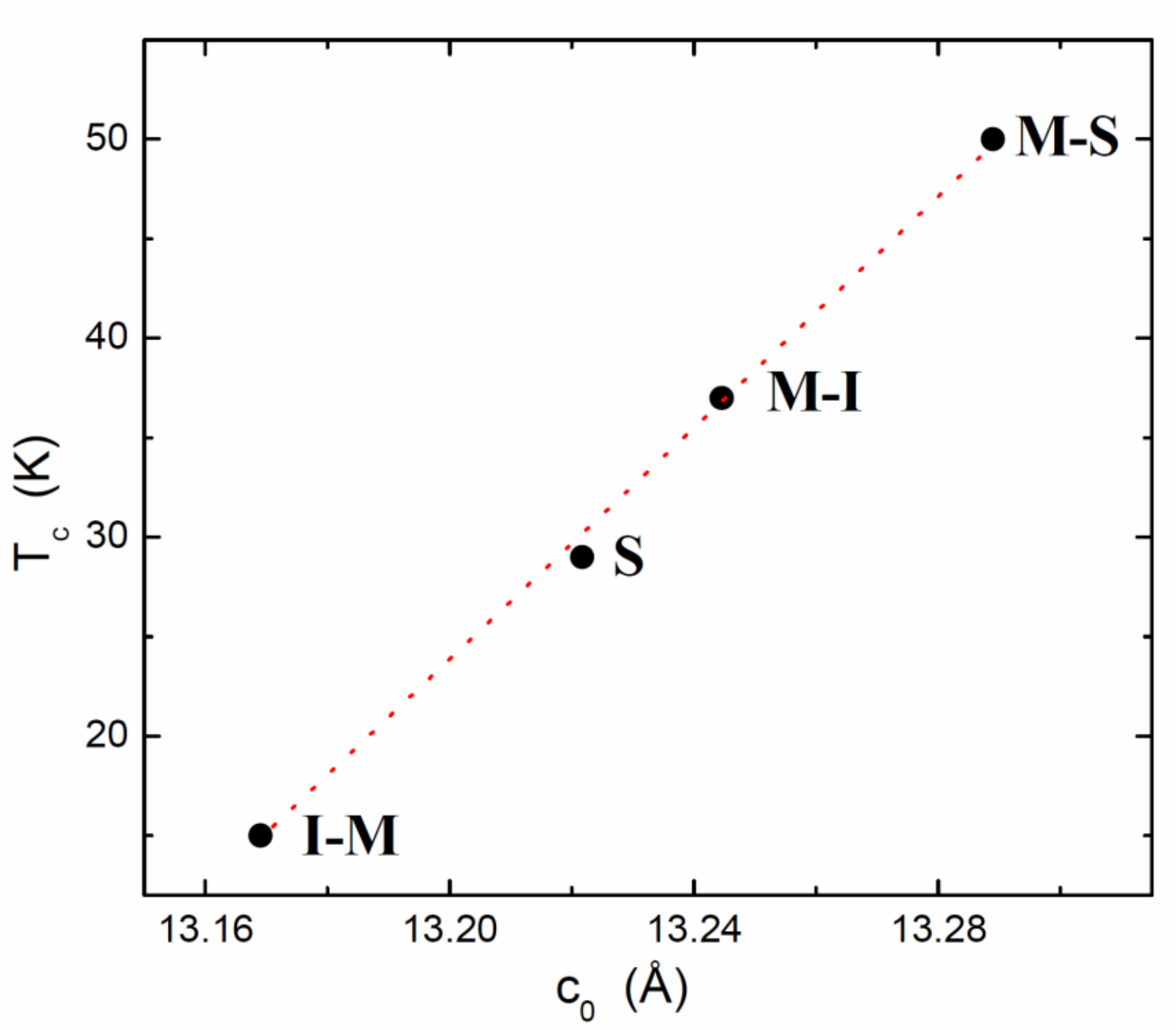

**Figure 15**. The relation of $T_c$ and the out-of-plane lattice constant ($c_0$) in $La_2CuO_{4+\delta}$ superconducting layers and in *I-M*, *M-I* and *M-S* bilayer structures. The red dashed line is a linear fit to the data. [144]

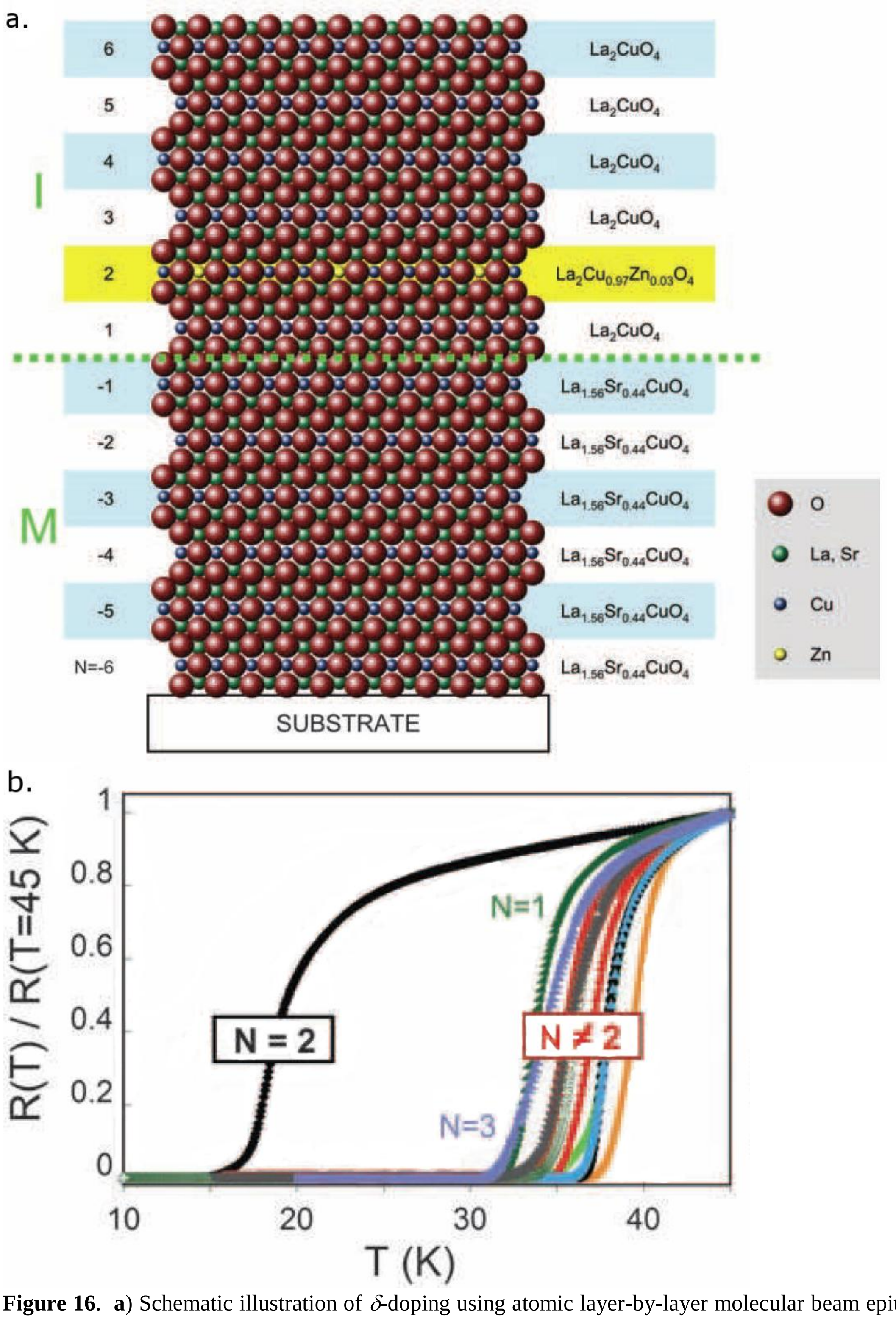

**Figure 16**. **a**) Schematic illustration of $\delta$-doping using atomic layer-by-layer molecular beam epitaxy (ALL-MBE). The model represents a six UC thick $La_{1.56}Sr_{0.44}CuO_4/La_2CuO_4$ (*M-I*) bilayer. The green dashed line indicates the position of the geometrical interface between the layer $N$ = -1 belonging to *M* and the layer $N$ = 1 belonging to *I*. The yellow layer ($N$ = 2 in the figure) highlights the monolayer where the Zn $\delta$-doping has been introduced. **b**) The effect of the Zn $\delta$-doping of the bilayer structures in the normalized resistivity, *R(T)/R(45K)*, as a function of temperature *T*. *N* indicates the position of the doped monolayer within the structure as indicated in panel a. [77]

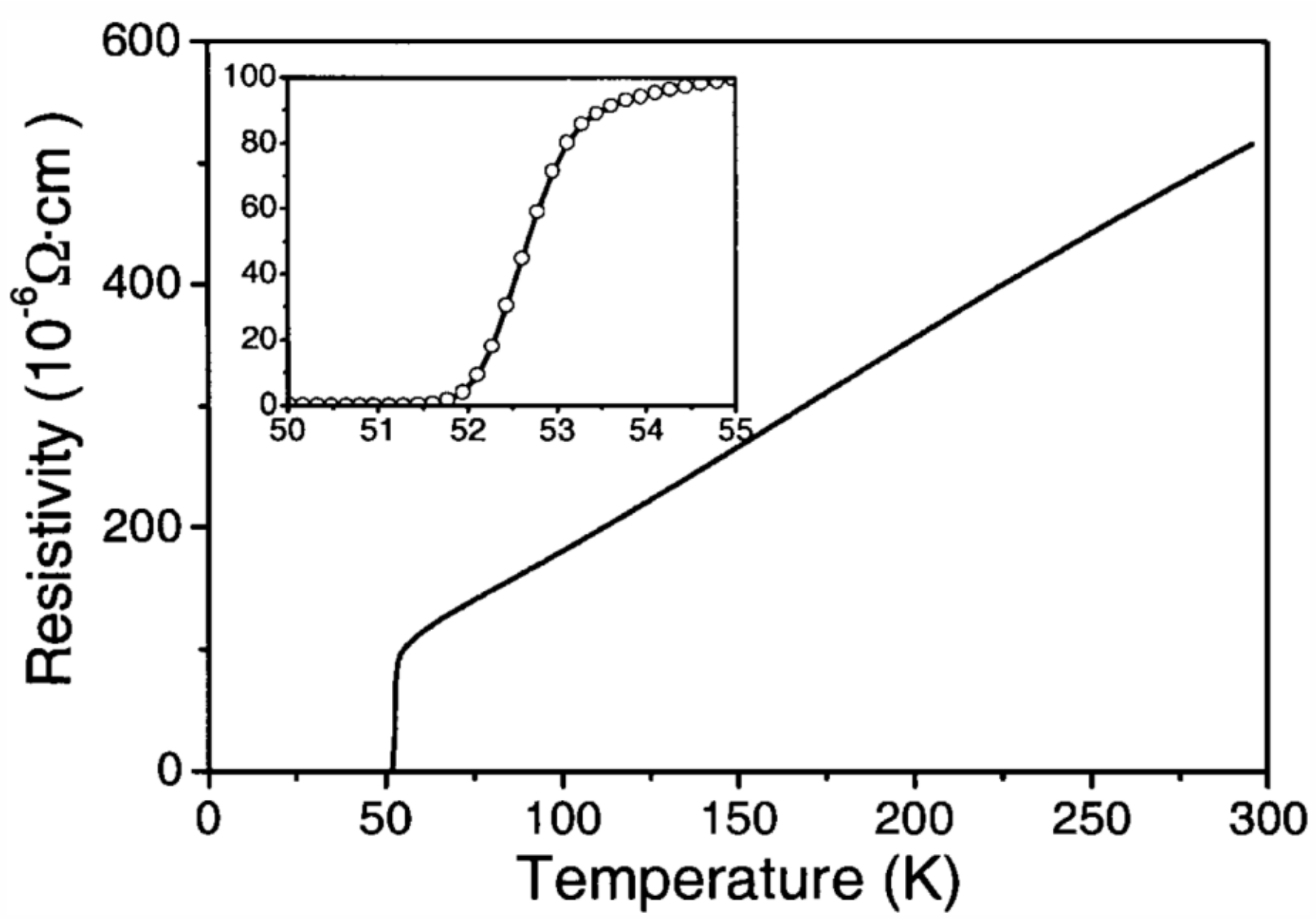

**Figure 17**. Resistivity as a function of temperature of a $La_{1.85}Sr_{0.15}CuO_4/La_2CuO_{4+\delta}$ (*S-S'*) heterostructure. [61]

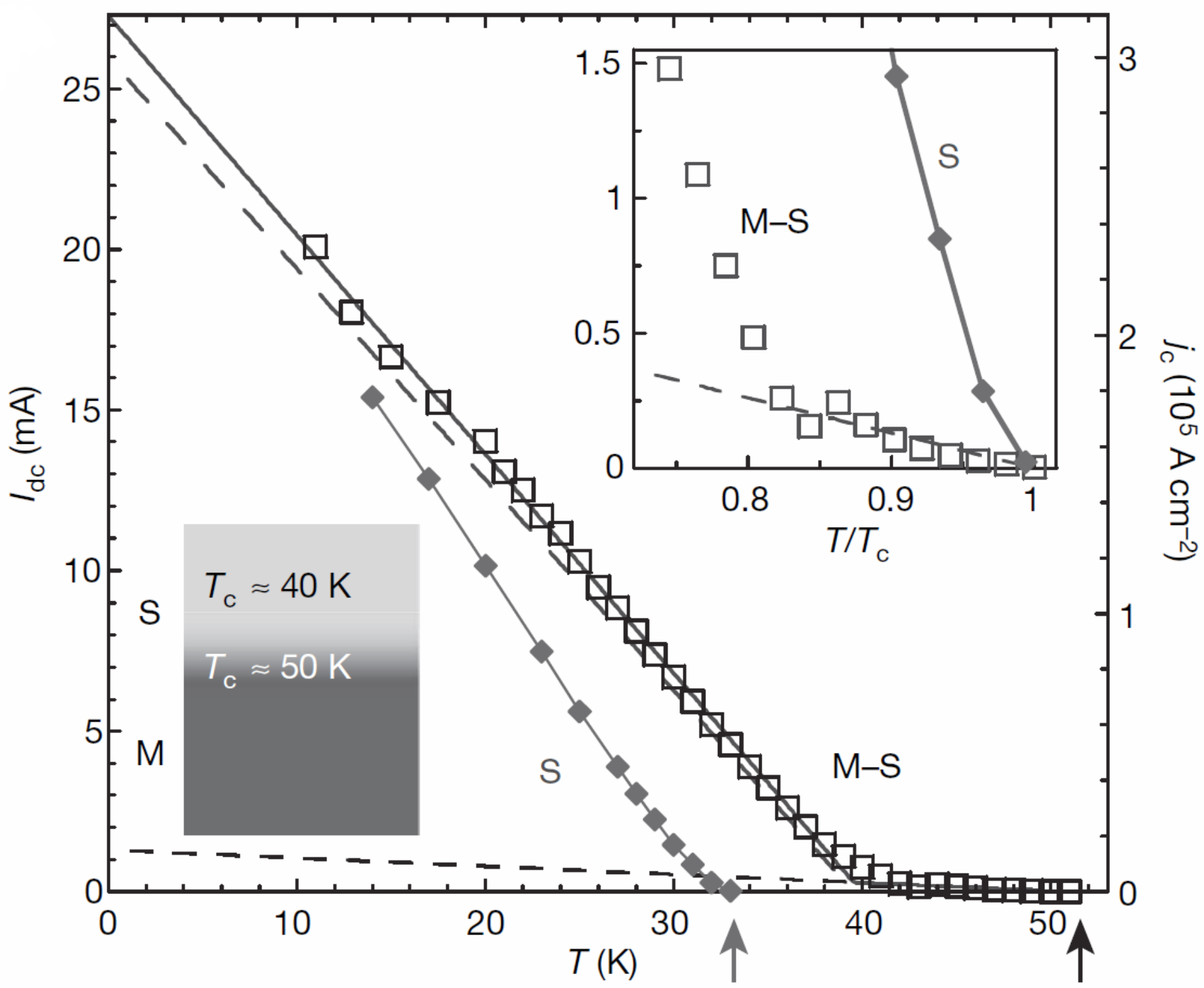

**Figure 18**. Temperature dependence of $I_{dc}$ for a $La_2CuO_{4+\delta}$ (*S*) film (solid diamonds) and a $La_{1.56}Sr_{0.44}CuO_4/La_2CuO_{4+\delta}$ (*M-S*) bilayer (open squares). The right scale shows the calculated peak value of the induced screening current density in superconducting films. Arrows denote the values $T_c$ = 33.2 K and 51.6 K for the *S* and *M-S* samples, respectively. The bilayer data can be well decomposed into two approximately linear contributions (dashed lines), corresponding to bulk and interface parts with $T_c \approx 40$ K and $T_c \approx 50$ K (lower left inset). Top right inset shows the same data in reduced temperature units, $T/T_c$. [60].

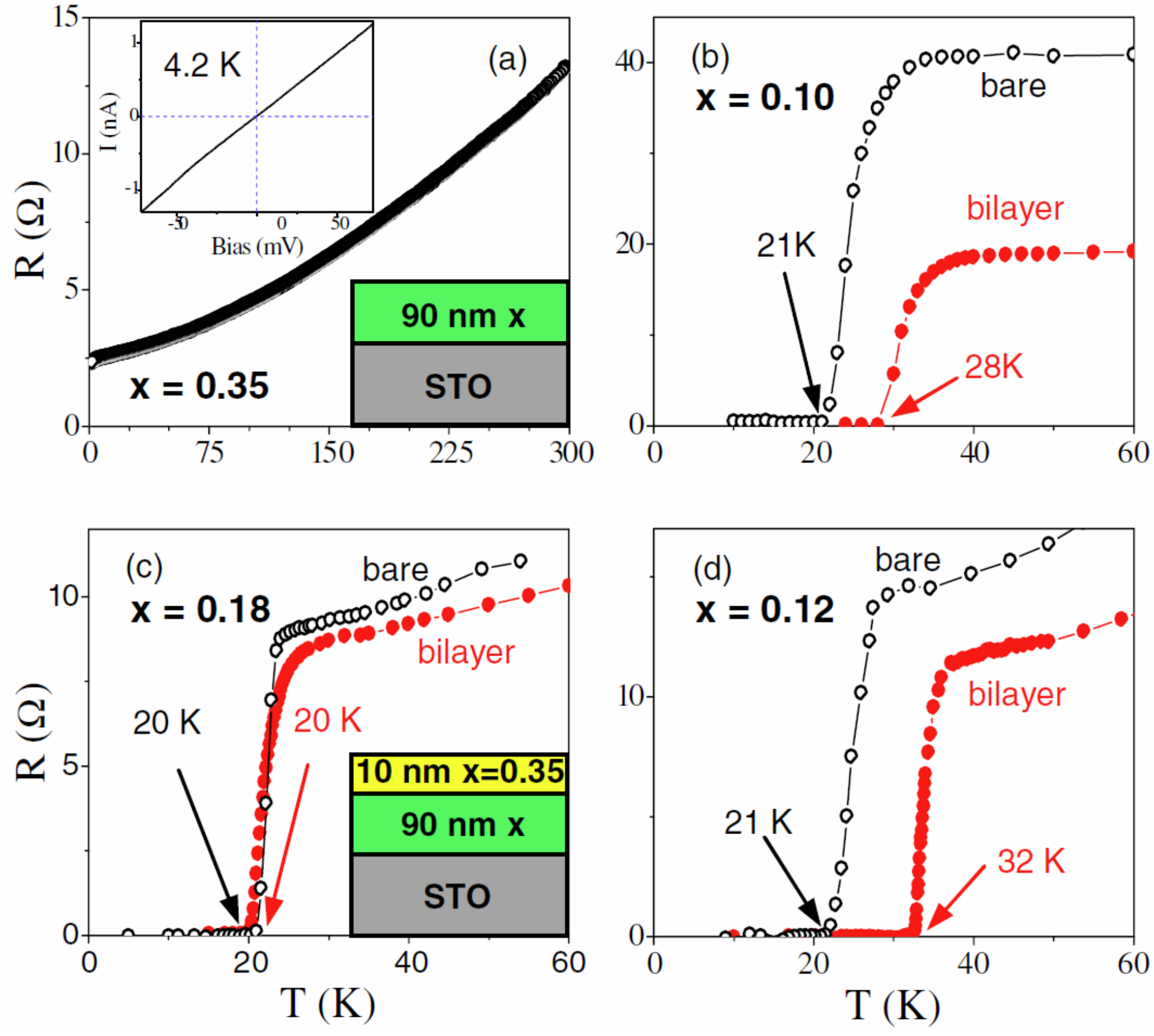

**Figure 19**. **a**) Resistance as a function of temperature of a bare $La_{1.65}Sr_{0.35}CuO_4$ film. The inset depicts the *I*-*V* tunneling characteristic of the same film taken by STM at 4.2 K. **b**)-**d**) Resistance as a function of temperature of the $La_{1.65}Sr_{0.35}CuO_4$-$La_{2-x}Sr_xCuO_4$ bilayers with $x$ = 0.1, 0.18 and 0.12 and of the corresponding bare films. The arrows mark the zero-resistance transition temperature. [91]

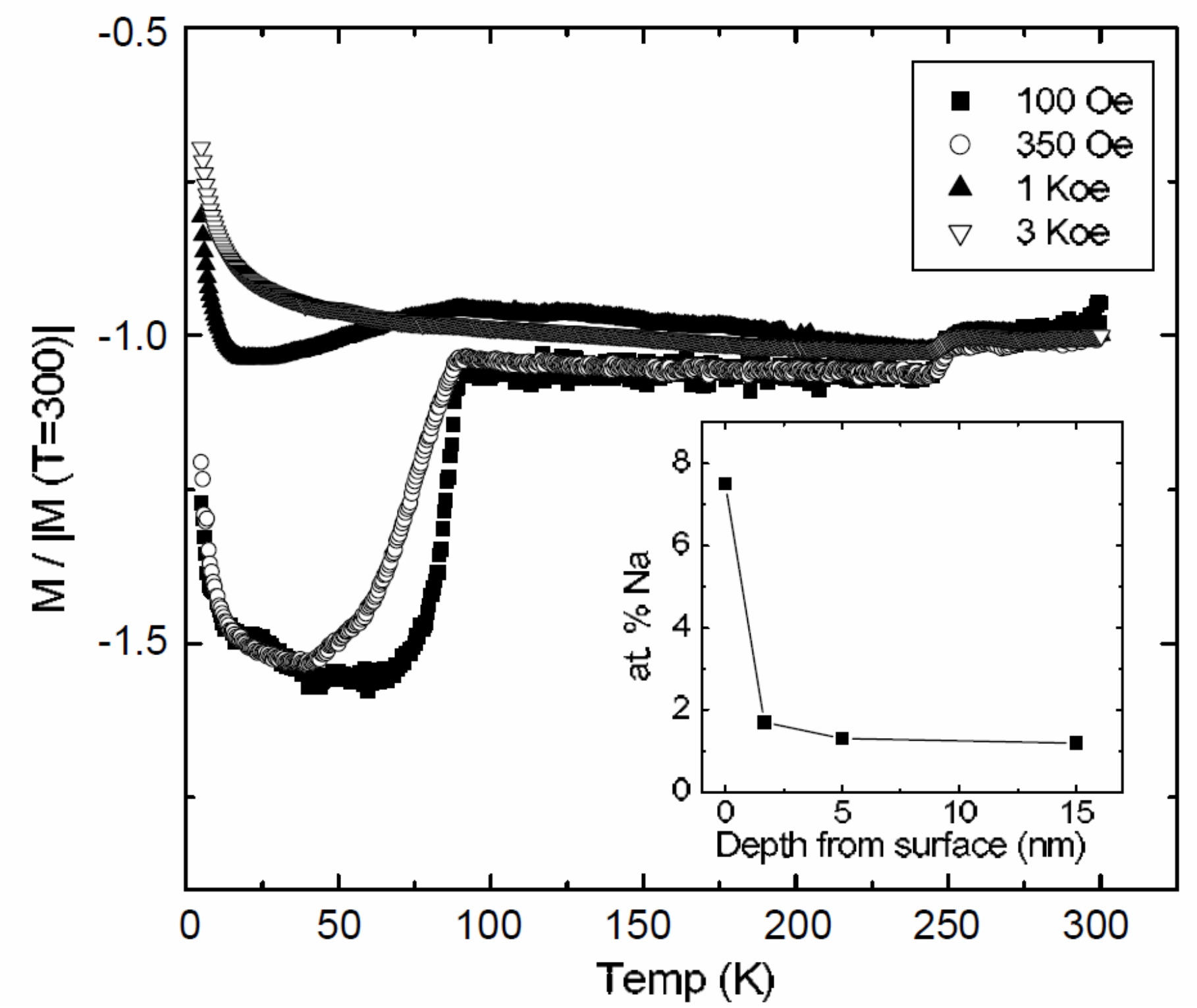

**Figure 20**. Zero-field cooled magnetization as a function of temperature of $WO_3$ crystals for various applied magnetic fields. Each curve is normalized to the absolute value of the magnetization at $T$ = 300 K. The XPS measurement of the Na content of the layer as a function of depth is shown in the inset. [102]

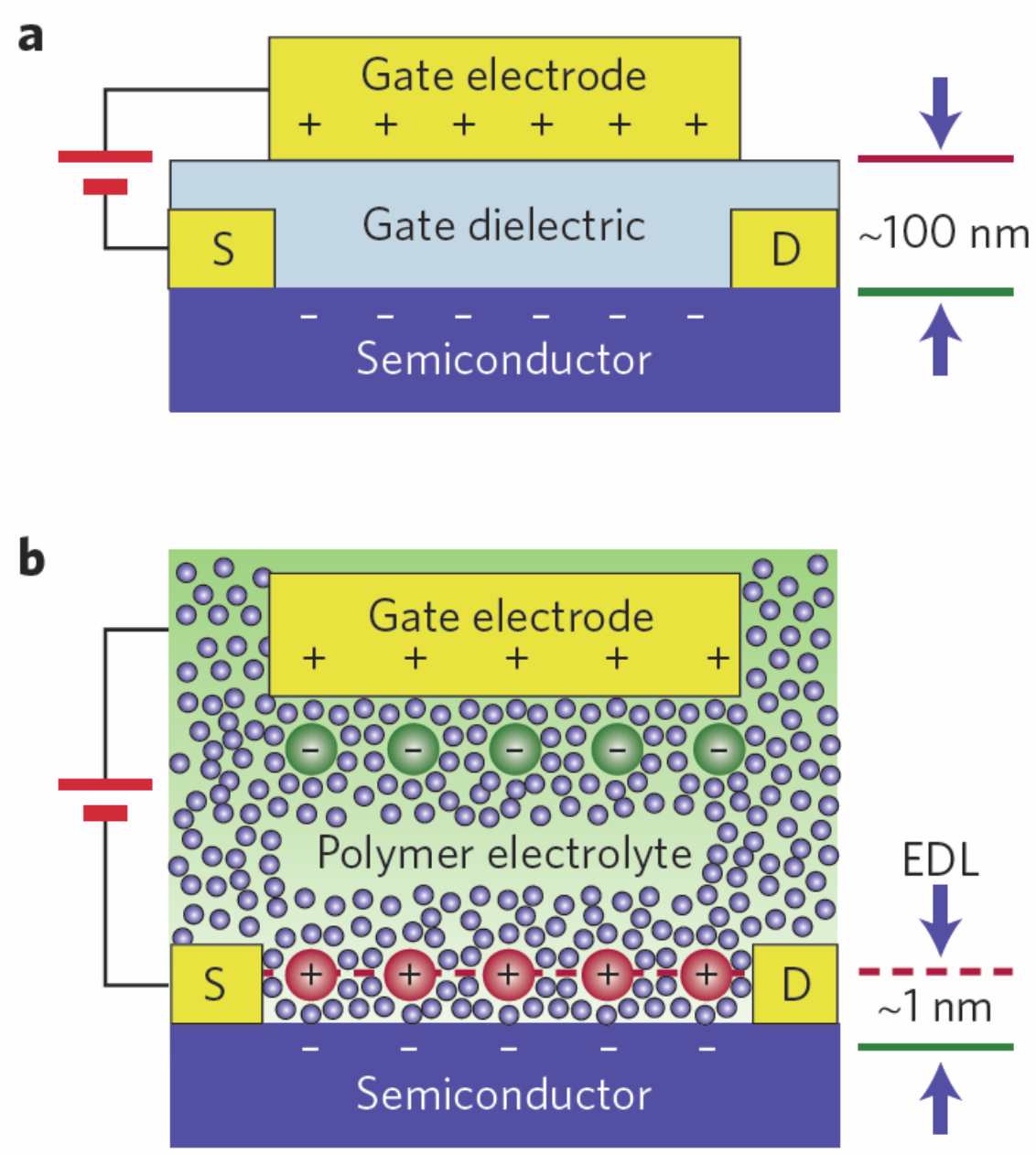

**Figure 21**. **a**) Conventional FET device incorporating thick (~ 100 nm) gate dielectric. The -/+ symbols are labels describing the charge accumulation/depletion layers on the surface of the semiconductor channel/gate electrode under an applied positive gate voltage. S and D denote the source and the drain, respectively. **b**) An electrolyte double-layer transistor (EDLT) device. [145]

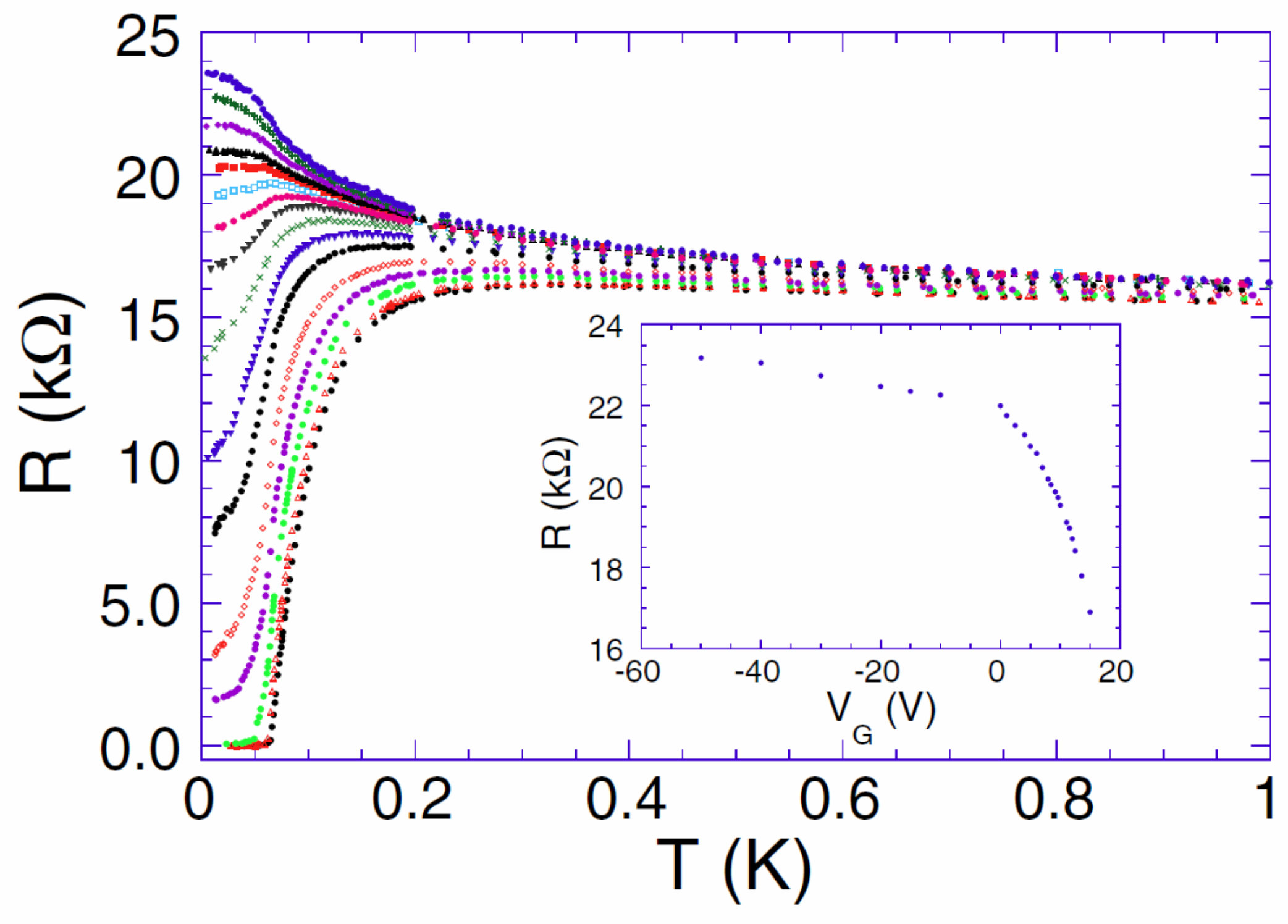

**Figure 22**. *R(T)* as a function of gate voltage for a 10.22 Å thick Bi film. From top to bottom the voltages are 0, 2.5, 5, 7, 8, 9.6, 11, 12.5, 14.5, 17, 19.5, 24, 28, 33, 38 and 42.5 V. The inset shows the resistance as a function of gate voltage for the same film at 65 mK. [115]

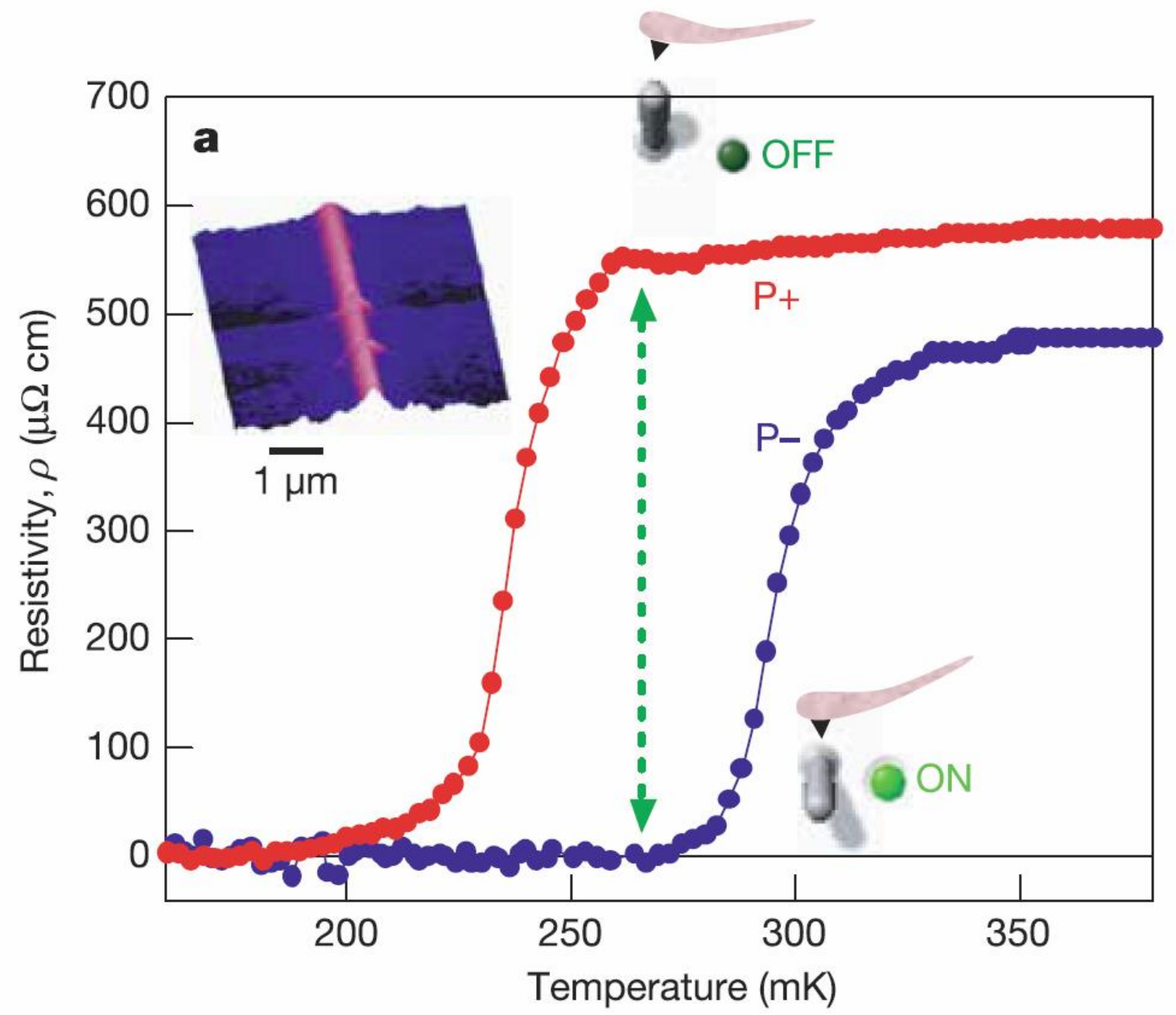

**Figure 23**. Resistivity as a function of temperature of Nb doped $SrTiO_3$. As indicated by the arrow and the associated insets, superconducting switching is observed at ~270 mK. The left inset shows the piezo-response image after poling a $P^+$ state line in the $P^-$ state background. [116]

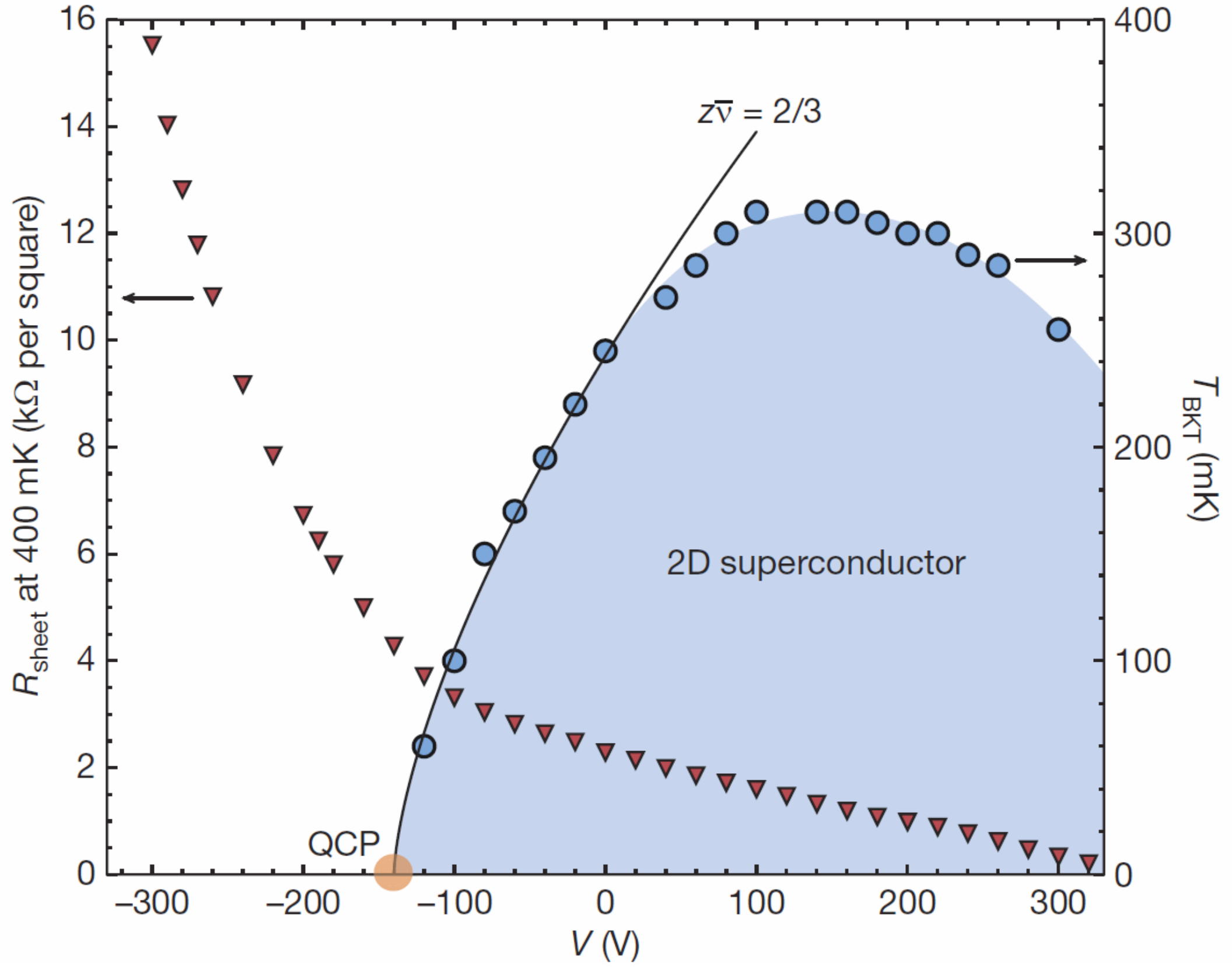

**Figure 24**. Electronic phase diagram of the $LaAlO_3/SrTiO_3$ interface. $T_c$ is plotted as a function of the gate voltage (right axis, blue dots), revealing the superconducting region of the phase diagram. The solid line describes the approach to the quantum critical point (QCP) using the scaling relation $T_{BKT} \sim (\delta V)^{zv}$ with $zv = 2/3$. The normal-state resistance measured at 400 mK is also represented as a function of gate voltage (left axis, red triangles). [117]

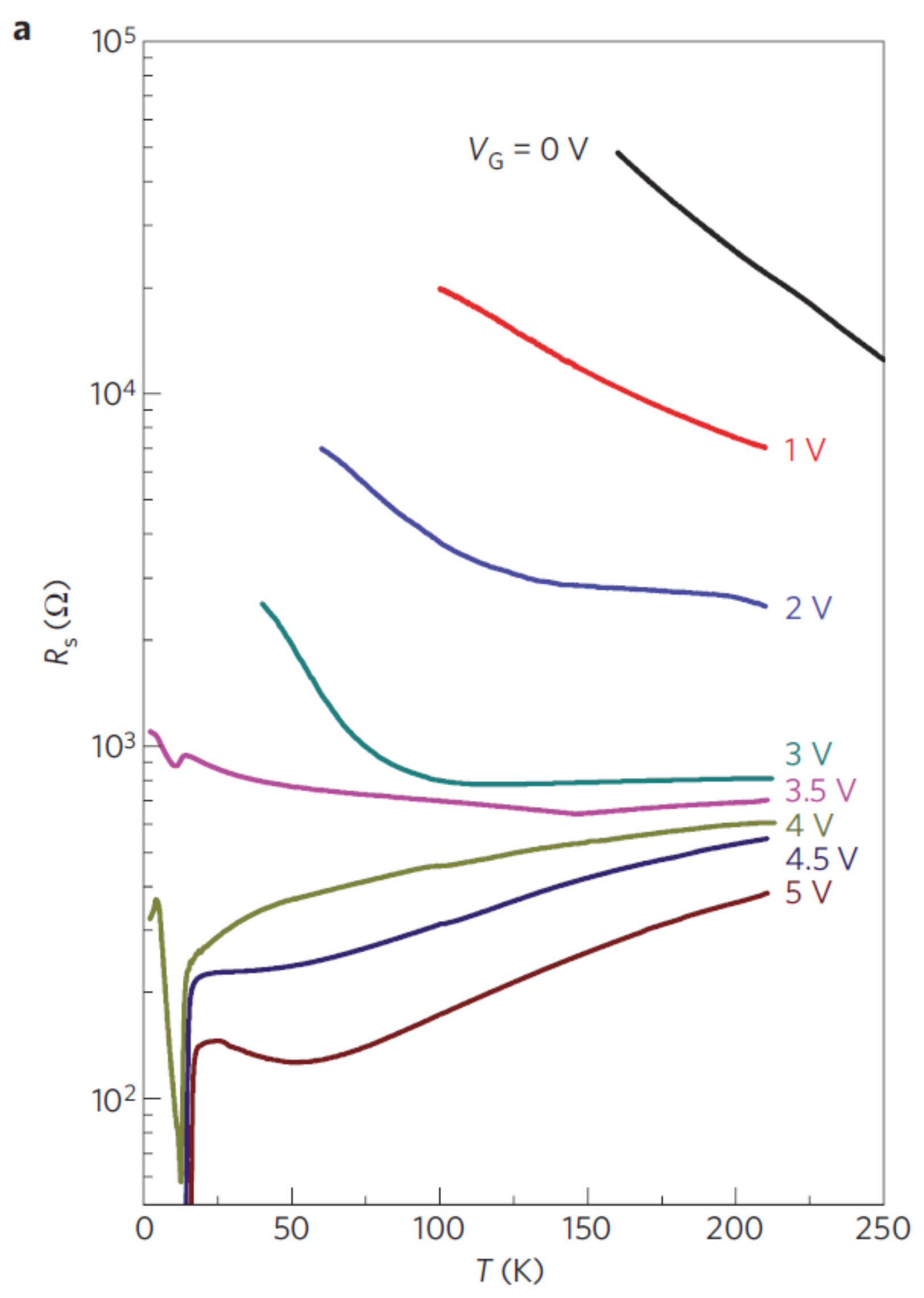

**Figure 25**. Temperature dependence of the channel sheet resistance at different gate voltages for a $ZrNCl/SrTiO_3$ bilayer. [146]

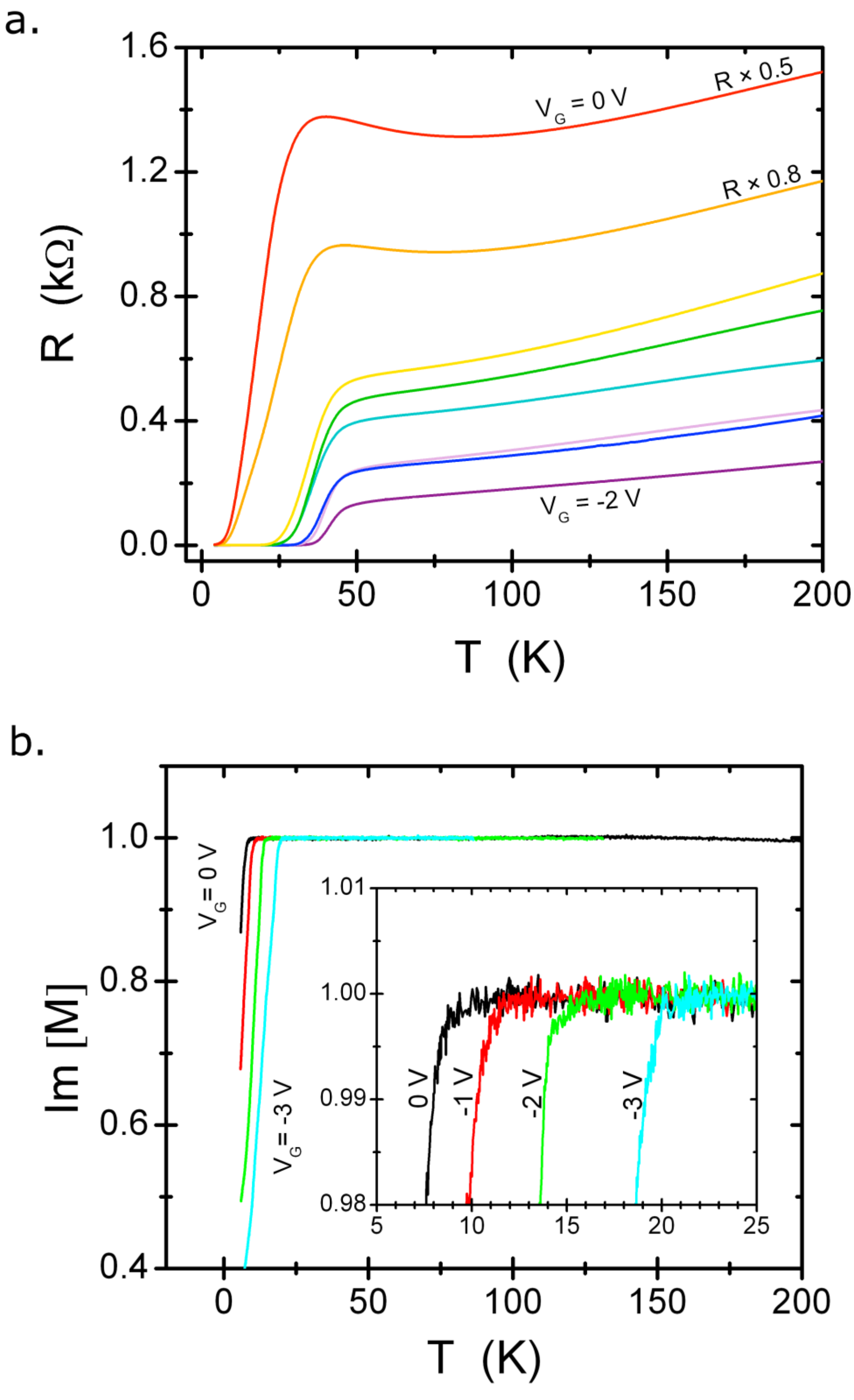

**Figure 26**. **a**) Resistance as a function of temperature in EDLT device containing 1 UC thick $La_{2-x}Sr_xCuO_4$ layer, for different gate voltages. **b**) Diamagnetic screening of magnetic field (measured by the mutual inductance technique) of $La_{2-x}Sr_xCuO_4$ layers at different gate voltages. [132]